\documentclass[review]{elsarticle}

\usepackage{lineno,hyperref}
\modulolinenumbers[5]

\journal{Annals of Physics}

\usepackage{lipsum}
\usepackage{natbib}
\usepackage{graphicx}
\usepackage{dcolumn}
\usepackage{bm}
\usepackage{dcolumn}   
\usepackage{bm}        
\usepackage{amssymb, amsmath}   
\usepackage{enumerate}
\usepackage{slashed}
\usepackage{simplewick}
\usepackage{comment}
\usepackage{color}
\usepackage{soul}
\usepackage{empheq}
\usepackage{appendix}
\usepackage[utf8]{inputenc}
\usepackage[english]{babel}
\usepackage{geometry}
\usepackage{marginnote}
\usepackage[normalem]{ulem}

\newcommand{\ii}{{\rm{i}}}

\newcommand{\vv}{{\rm {v}}}

\newcommand{\nn}{\nonumber}

\newcommand{\eq}[1]{(\ref{#1})}
\renewcommand{\>}{\rangle}
\newcommand{\la}{\label}
\newcommand{\ba}{\begin{align}}
\newcommand{\ee}{\end{equation}}
\newcommand{\be}{\begin{equation}}

\def\12{\frac{1}{2}}
\newcommand{\p}{\partial}
\newcommand{\en}{\end{align}}

\newcommand{\<}{\langle}

\usepackage{color}

\def\red{\color{red}}

\begin{document}

\begin{frontmatter}

\title{Geometry  of Quantum Hall States: \\Gravitational Anomaly
and Transport Coefficients }
\author[1]{  Tankut Can } 
\address[1]{Simons Center for Geometry and Physics, Stony Brook University, Stony Brook, NY 11794, USA}
 \cortext[mycorrespondingauthor]{Corresponding author}
\ead{tcan@scgp.stonybrook.edu}
\author[2]{Michael Laskin }
\author[2]{Paul B. Wiegmann }
\address[2]{Department of Physics, University of Chicago, 929 57th St, Chicago, IL 60637, USA}

\date{\today}

\begin{abstract}\noindent
We show that  universal  transport coefficients of  the fractional
quantum Hall effect (FQHE)  can be understood as a response to  variations
of  spatial geometry. Some transport properties are essentially governed by the gravitational anomaly. We develop a general method to compute  correlation functions of FQH states in a curved space, where local transformation properties of  these states are examined through local  geometric variations. We introduce the notion of a generating functional and relate it to geometric invariant functionals recently studied in geometry. We develop two complementary methods to study the geometry of the FQHE. One method is based on iterating a Ward identity, while the other is based on a field theoretical formulation of the FQHE through a path integral formalism.  
  \end{abstract}



\end{frontmatter}


\section{Introduction}

Most of the universal features of the quantum Hall effect (QHE) are understood through transport
properties of the electron gas at an energy scale much less than the activation energy (the cyclotron energy in the integer quantum Hall (IQH)
case, the Coulomb energy in the fractional quantum Hall (FQH) case). These coefficients, such as the Hall conductance, are measured in low-energy excited edge states since all the bulk states are gapped. 

However, it was observed long ago that the same  transport coefficients can be extracted from the ground state on a compact manifold when the QH droplet covers the surface entirely. This suggests a correspondence between QH edge states in flat space and bulk states on a closed manifold.  An example is the St\v reda formula \cite{Streda1982}, which states that
the Hall conductance  along the edge of a flat sample is equal to the variation of the g state bulk electronic density with respect to the magnetic field 
\begin{align}\la{0}\sigma_{H} = \frac{e\delta \rho}{\delta B} ,
\end{align}
where $ \rho$ is the particle density, $e$ is the electron charge, and \(B\) is the magnetic field.

We show that this correspondence is significantly deeper. Namely, all dissipation-free  transport coefficients at low energies of the QHE can be understood as the geometric response of the g state on a   closed manifold. This also holds for all other  topological states.  To be precise, ``geometric" means a response to scalar curvature. We also refer to it as the gravitational response, since geometry is encoded in a metric.

Like the Hall conductance, other transport coefficients are also related to gravitational response of  QH states. These coefficients reflect  anomalies of the transformation of quantum states, and for that reason  must be quantized on QH plateaus like the Hall conductance. The Hall conductance is associated to the gauge anomaly, while other transport coefficients, such as the odd (or anomalous) viscosity and thermal conductivity are associated with mixed gauge and gravitational anomalies, respectively. While the Hall conductance depends only on the filling fraction, other  coefficients reveal additional universal features of FQH states.  Although difficult to measure experimentally, these features are important for understanding the theory of the QHE.




In this paper we study QH states on Riemannian manifolds to develop  a  physical understanding of the QHE through geometry. We focus especially on FQH states described by Laughlin's wave function \cite{Laughlin1983}.  Since ground states on curved surfaces  provide a complete
description of the QHE on a flat background, this is a useful method for understanding the former, experimentally observable states.  More importantly, this approach implies
geometric reasons for quantization and uncovers universal features of the
QHE that are inaccessible from calculations in flat space.

Our approach is based on the extension of the  Laughlin wave function to a Riemann surface.   We will argue that  the  generalization which  preserves the holomorphic structure of the Laughlin-Jastrow factor is unique. For surfaces with constant negative curvature, it was carried out in Ref. \cite{Iengo1994} following  Refs. \cite{Haldane1983,HaldaneRezayi} where the Laughlin wave function was written for a sphere and the torus. The relation between the total flux and the number of particles in FQH states without boundary was established in these papers. The wave function of the completely filled  lowest Landau level (i.e. the integer quantum Hall effect) on surfaces with arbitrary curvature first appeared in \cite{Klevtsov2013}. In  Sec. \ref{fqh_model_wf} we present the arguments which determine the Laughlin wave function (the fractional quantum Hall effect)  on an arbitrary Riemann surface. We focus on genus zero surfaces.  

Naturally, it is not expected that the FQH ground state of an interacting electron system  has exactly the form of the Laughlin wave function, especially for arbitrary background fields. However, we argue that if one is concerned only with the universal features  of the FQHE, the model wave function we use can serve as a lamppost. It captures the universal features of the phenomena. For further discussion of this and related matters, we refer to recent papers \cite{Haldane2009, Golkar2013}. Comparison with {\it ab initio} numerical studies would be desirable in this respect. 

Adopting this perspective, the authors recently argued in Ref.\cite{CLW} that
electromagnetic transport in the QHE is directly connected to the response
of the ground state under variations of its spatial geometry and showed that
the gravitational response is controlled by a gravitational anomaly, a new
phenomenon missed in previous studies of the QHE. The aim of this paper is
to provide the details of the argument outlined in \cite{CLW}.

 The importance of 
geometry (as opposed to topology) in the QHE was emphasized in an early seminal study of adiabatic transport in moduli spaces (on example of a torus)  by Avron, Seiler, and Zograf
\cite{Seiler1995} and L\'evay \cite{Levay,Levay1997} where the notion of  the anomalous (or odd) viscosity was introduced. That notion has been extended   to FQH model wave functions more recently in Refs. \cite{Tokatly2009, Read2009, Read2011}. The geometric
response of the QHE was first studied in \cite{Wen1992a,Frohlich1992}. Response of the completely filled lowest Landau level (the integer quantum Hall effect) to slowly varying electromagnetic and gravitational fields was recently addressed in \cite{Abanov2013}. The study   of  the QHE  coupled to  geometry  in the framework of `effective field theory' and  in a hydrodynamic approach was carried out  in \cite{ Abanov_solo, SonNC, GromovAbanov, fradkin2014,framinganomaly}  and \cite{PW12_arxiv, PW13, Wiegmann2013}. The results of this paper, based on the explicit form of the wave function, are complementary to and can inform the construction of the `effective' and hydrodynamic descriptions. Other works on the subject of the Laughlin  wave function in various backgrounds include
\cite{ ABWZ, Zabrodin2006}. 
 
 Results for the electronic density in the lowest Landau level (IQHE)  on Riemann surfaces were obtained by computing the gradient expansion in curvature of the Bergman
kernel in \cite{Klevtsov2013, Douglas2010}. Our work extends these results to the FQH
 case. The analysis
of the FQH case requires special analytic methods developed in \cite{Zabrodin2006}. We employ them here.

We note that  the mathematical objects that appeared in the theory
of the IQHE  are also significant in the context of K\"ahler
geometry, where  similar objects were introduced 
in the framework of the Yau-Tian-Donaldson program   \cite{donaldson2001scalar,donaldson2005scalar,berman2011kahler,berman2013thermodynamical,PhongSturm, Tian2000}. This connection was first appreciated and studied in great detail in Ref.\cite{Klevtsov2013}.

Geometry enters the IQHE naturally through the effect's global features, such as the degeneracy of the lowest Landau level (LLL). At  integer filling, the Riemann-Roch theorem implies that the number of admissible  states \(N_1\)
in the LLL on a closed surface is shifted  from  the number of flux quanta \(N_\phi\)  piercing the surface by half of the Euler characteristic of the surface  \(N_1=N_\phi+\chi/2\). Physically, this means that  the number of allowed states is equal to the number of flux quanta  only for surfaces  topologically equivalent to a torus.  For example, on surfaces of the conformal class of a sphere, $\chi=2$ and the number of particles exceeds the number of flux quanta by 1. Similarly, for the conformal class of the a two-torus, $\chi=-2$ and the number of allowed states is 1 less than the number of flux quanta.
The relation between the admissible number of particles on the lowest Landau
level and the number of the flux quanta can be generalized to FQH states. In particular, Laughlin states on the sphere and the torus with constant curvature have been constructed in early papers \cite{Haldane1983,HaldaneRezayi}, and later extended to higher genus constant negative  curvature surfaces \cite{Iengo1994}. It was observed that the relation is  modified by
the filling fraction \(\nu\) to \(N_1=\nu N_\phi+\chi/2\). 

General  FQH states beyond Laughlin states feature new characteristics, such as the shift $\mathcal S$ introduced in \cite{Wen1992a}, and the spin $s$ of particles explained in Sec. \ref{IQHresults} below. In this case the maximal  admissible number of particles in the ground state is 
\begin{align}
N_\nu & = \nu \Big (N_{\phi}  +  (\mathcal{S} -2s)  \frac{\chi}{2} \Big )\la{shift-spin}.
\end{align}

For the Laughlin states, $\mathcal{S}=\nu^{-1}$. For Pfaffian states \(\mathcal{S}=\nu^{-1} + 1 \)  \cite{Read99}.  A more detailed definition of the shift  is given in the Sec.\ref{221}.

The global relation  \eq{shift-spin} can be expressed locally. We define the density such that the number of particles in a volume element  in  a given state
is   $\rho d V$.  The expectation value of the density in the ground state is then 
\begin{align}\la{1}\langle\rho\rangle=\nu \Big (\frac B {\Phi_0} +(\mathcal{S}-2s)\frac R{8\pi} \Big ) ,
\end{align}
where $R$ is the scalar curvature, and $\Phi_{0} = h/e$ is the flux quantum, and $h$ is the  Planck constant. This formula appears in different but equivalent forms in  \cite{Wen1992a, Frohlich1992}.  The Gauss-Bonnet theorem \(\int R dV=4\pi\chi\) ensures that \eq{1} yields \eq{shift-spin} after integration  over the entire surface.

Equation \eq{1}  shows that particles accumulate on areas of larger curvature.  It also suggests that curvature acts as an effective  magnetic field. However, Eq.\eq{1} is not exact but an approximation to the density, which has higher order  gradient corrections in curvature and  magnetic field.  These corrections
break the apparent  similarity between the scalar curvature and magnetic field as Eq.
\eq{1} may suggest. 
For a constant magnetic field, the corrections appear at higher orders in $l^2$, where $l = \sqrt{\hbar/eB}$ is the magnetic length.  The next order term in this approximation is governed by the gravitational anomaly, evident in the transport properties of the QHE. We compute it in this paper. Note that the earliest indication of the gravitational anomaly for the Laughlin states appeared in an early paper by Jancovici \cite{Jancovici}, which focused on the mathematically identical system of a 2D Coulomb plasma, also known as a Dyson gas.

We focus on the Laughlin state \cite{Laughlin1983} on a deformed sphere and  discuss others briefly, although the current understanding of the gravitational anomaly for other states is incomplete. While our computations assume a surface which belongs to the conformal class of the sphere, some of our formulas are local  and therefore applicable to surfaces of arbitrary genus. Also, in most of the paper we assume 
that gradients of the curvature are small compared with the magnetic length. For an inhomogeneous magnetic field, we assume small variations over a large constant background.

The paper is organized as follows.
We start (Sec.\,\ref{Sec2}) by describing free electrons in the lowest Landau level on a curved
surface. Then we extend the results to the FQHE in a curved space. In
this Section we emphasize the role of M\"obius transformations in the FQHE. In Sec. \ref{Sec3} we introduce the central object of the study,
the generating functional, and show how to obtain the linear response
of the system under various deformations. In Sec. \ref{Sec4} we connect some transport coefficients -  the Hall
conductance, the odd (or anomalous) viscosity and the static structure factor to the response of electronic  density to variations of curvature. {  In Sec. \ref{hallviscosity_structurefactor} we comment briefly on a connection to some results in \cite{Haldane2009, Golkar2013}.}
In Sec. \ref{Sec6} we obtain the Ward identity for the FQH states. In Sec. \ref{UV_reg} we complement the Ward Identity by the short distance regularization
 of various operators. This converts the Ward identity into an equation, which
we  iterate  to obtain  the gradient expansion of the transport coefficients in Sec. \ref{Sec7}. In Sec. \ref{Sec9}
we discuss various forms of the generating functional and link it to geometric invariant functionals 
studied in modern geometry. In Sec.\ref{E} we apply the results for the generating functional to extend the results to the case of a non-homogeneous magnetic field. In Sec. \ref{Sec10}
we discuss a conjecture about extending our results to other FQH states. In Sec.
\ref{4} we present an alternative calculation by virtue of a formulation of the Laughlin states
in terms of the Gaussian Free Field. This formulation
yields  a straightforward and transparent  derivation of the generating functional
through the gravitational anomaly.

The appendices are a collection of observations tangent to the main content of the paper. Some of them are the addenda to calculations and references, while others are subjects of inquiry that are interesting in their own right. In the appendices we discuss more general FQH states (such as the parafermionic state), we demonstrate  how the techniques developed in the paper reproduce known results about the charge, spin and statistics of the
quasi-holes, relate the odd viscosity to the response to curvature, and further discuss geometric invariant functionals and transformation formulas.  Finally we relate our results to the thermodynamics of the 2D Coulomb plasma. In this interpretation the gravitational anomaly appears as a gradient  correction to the Boltzmann entropy.

\subsection{Main Results}

 We list two results of the paper.  One is the expansion of the particle density of FQH states on a closed surface in covariant derivatives and powers of intrinsic curvature. In a uniform magnetic field $B$, it is 
\begin{align} \la{rho}
\langle \rho \rangle  = \frac {\nu}{2\pi l^{2}} +  
 \frac{\nu (\mathcal{S}-2s)} {8 \pi}R+\frac{ b}{8 \pi} l^{2} \Delta_{g} R+O(l^4),
\end{align}
where $l$ is the magnetic length, \(\Delta_g\) is the Laplace-Beltrami operator, and \(R\) is scalar curvature. This can be converted to an expansion in the total magnetic flux in units of the flux quanta $N_{\phi} = B V/\Phi_{0}$ through the relation \(l^{2}=  V/(2\pi N_\phi)\). The large $N_{\phi}$ expansion is equivalent to the small $l$ expansion. 

We present this formula in an equivalent form by virtue of \eq{shift-spin}
\begin{align} \la{rho1}
\langle \rho \rangle  =  \frac {N}V +  
\nu \frac{\mathcal{S}} {8 \pi}(R-\bar R)+\frac{ b}{8 \pi}l^2 \Delta_{g} R+O(l^4),
\end{align} 
Here $\bar R=\frac 1V\int R dV
=4\pi\chi/V$ is the mean curvature.

The coefficient $b$ is given by 
\begin{align}
b = -\frac{\kappa}{12} + \frac{\nu (\mathcal{S}-2s)(2 - \mathcal{S})}{4},
\end{align}
where the number {  $\kappa $} is controlled by the gravitational anomaly, which we define and discuss at length in the text.  It has been suggested that this coefficient fixes the thermal response \cite{GromovAbanov}, although a direct derivation is lacking.

The  structure of the formula holds for any FQH state. The data which characterize a state are the filling fraction $\nu$, the shift $\mathcal{S}$, and the gravitational anomaly {  $\kappa$.} We compute it for  the Laughlin states, for which the shift $\mathcal{S} = 1/\nu$ is an integer, and find   
\begin{align}\la{1.7}
{  \kappa = -1,} \quad b=\frac 1{12} + \frac 1{2\nu}(1 - 2 \nu s ) \left(\nu - \frac 12 \right).
\end{align}
For other FQH states the value of {  $\kappa$} is discussed in Sec.\ref{Sec10}. 
 We comment that at $\mathcal{S}=2s$ or $\mathcal{S}= 2$, the gravitational anomaly is the sole source of the coefficient $b$. In  this case it becomes  $-\kappa/ 12$. In particular, this occurs for Dirac particles ($s=1/2, \nu=1$), for the Laughlin Bose state $\nu=1/2$ or bosonic Pfaffian state ($\nu = 1, \mathcal{S} = 2$). 

The  higher order terms in \(l^2\) are currently inaccessible by our methods, and require a more detailed understanding of the short distance physics.  We will elaborate on this issue in Sec. \ref{UV_reg}. 

The integral of the first two terms in \eq{rho} yields the  number of  particles for the FQH state via the Gauss-Bonnet theorem.
The first term is related to the Hall conductance, the second term is the
result of the ``mixed anomaly". It describes the response of   electric
 charge or current to  a deformation of geometry. This term  is closely related to the anomalous
viscosity (Sec.4).  The third term is a combination of the mixed and gravitational anomalies. The gravitational anomaly appears as the \(1/12\) term
in  formula \eq{1.7}.

Another result is the computation of the static structure factor (defined in \eq{3.11})
in the long wavelength limit.  For the Laughlin state
\begin{align}\la{1.8}S(k)=\frac{k^2}{2}\left[1-\frac{k^2}{2\nu}\left(\nu-\frac{1}{2}\right)+\frac{k^4}{4\nu}\left(\nu-\frac{1}{2}-b \right)+O(k^6)\right],\quad l=1.\end{align}
where $b$ is given by \eq{1.7}. The gravitational anomaly appears at order $k^{6}$. This term was first found in \cite{Kalinay2000} through a diagrammatic approach.
Interestingly, it vanishes for the \(\nu=1/3\)  state. We explain the
significance of the gravitational anomaly in the body of the paper but one comment  is in order here. The bosonic
Laughlin state \(\nu=1/2\) describes the states of vortices in the superfluid.
Classically, this system is conformally invariant.  If the state is conformally invariant
 in the quantum limit, then it will be evident in the structure factor, which reads \(S(k)=k^2/2\) at all \(k\). 
This symmetry is violated by the  gravitational anomaly.  In this
case \(b=1/12\) and the gravitational anomaly  shifts $S(k)$ from the conformally invariant value \begin{align}
S(k)=\frac{k^2}{2}\left(1-\frac{k^4}{24}+O(k^6)\right),\quad \nu=1/2.
\end{align}

 Other results  include the relation between the gradient expansion of the
Hall conductance and the gravitational anomaly, and computation of the  generating functional. 
 
The generating functional  $\mathcal{Z}[g]$ (see  \eq{gen_fun} for the definition) depends on the geometry of the manifold and the QH state. It encodes all universal properties of the QH states. The structure of the large $N_{\phi}$ expansion of the generating functional has the form 
\begin{align}\nn
  \log\frac {{\mathcal Z}[g]}{{\mathcal Z}[g_0]}=p_2 N_{\phi}^{2}
A^{(2)}[g,g_0]+p_1 N_{\phi}A^{(1)}[g,g_0]+p_0A^{(0)}&[g,g_0]+\sum _{k\geq 1}p_{-k}N_{\phi}^{-k}A^{(-k)}[g,g_0].
\end{align}
Here $g$  is the metric of the surface of interest and $g_0$ is the   metric of a reference  surface within the same  cohomology class.  For practical purposes the reference surface could be chosen to be a uniform  sphere. 

The numerical coefficients {  $p_k = p_{k}(\nu, \mathcal{S}, \kappa)$} are functions of the filling fraction, the shift, and the gravitational anomaly, while the functionals $A^{(k)}[g,g_0]$ are invariants of the manifold and do not depend on the state we elect.  Derived in Section \ref{9.2} and in the \ref{largeN},  the geometric functionals  read 
  \begin{align}
 A^{(2)}[g,g_0] &= \frac{1}{4 V^{2}} \int K dV - K_{0} dV_{0},\\
 A^{(1)}[g,g_0] & = \frac{2}{V} \int \log \sqrt{g}\, dV - \log \sqrt{g_{0}} \, dV_{0},\\
 A^{(0)}[g,g_0]& = \int R \log\sqrt{g}\, dV - R_{0} \log \sqrt{g_{0}} dV_{0}.
 \end{align}
Here $K$ is the K\"ahler potential (defined below). 

The first two geometric  functionals recently came to a prominence in K\"ahler geometry \cite{donaldson2001scalar}-
\cite{Tian2000} and also in the context of  2D quantum gravity \cite{Ferrari2012a}. The third functional is Polyakov's Liouville action \cite{Polyakov1987} reflecting the gravitational anomaly.
They are non-local functionals of the curvature. In contrast, the functionals $A^{(-k)}[g]\quad k\geq 1$ are local functionals of the curvature and covariant derivatives of the curvature.  We argue  that the three non-local  functionals encode  all of the universal properties of the FQHE and the first three coefficients $p_2,p_1,p_0$ in the above expansion  are  quantized on FQH plateaus just like the Hall conductance. We argue that all transport properties of FQH states are encoded by these coefficients. 
 We compute  them  for the Laughlin states, and outline a conjecture in Sec. \ref{Sec10} that this holds for general FQH states
  \begin{align}\la{10.2}
&p_2=- 2 \pi \nu,\\
&p_1=\frac{\nu  (\mathcal{S}-2s)}{4},\\
& {  p_0= \frac{1}{8\pi} \left( \frac{\nu  (\mathcal{S}-2s)}{2} - b\right) = \frac{1}{8\pi} \left( \frac{\nu (\mathcal{S}-2s)^{2}}{4} + \frac{\kappa}{12}\right). }
\end{align}

 The generating functional for arbitrary $\beta$ and spin $s$ was also derived recently in Ref.\cite{KlevtsovGFF} by an approach complementary to ours, discussed in Sec.\ref{14SS}. Our results for the generating functional and the particle density at $\beta = 1$ reproduce those of Ref.\cite{Klevtsov2013}, obtained by alternative methods.


\subsection{Notation}

We work in local complex coordinates $z = x +  \ii y$ and $\bar{z} = x - \ii y$ in which the metric is diagonal \eq{d} ($x$ and $y$ are known as isothermal coordinates) and denote a point on a surface as $\xi=(x,y)$. The volume element in these coordinates is $dV=\sqrt{g} dxdy=\sqrt gd^2\xi$. 

We  use the K\"ahler potential defined in \eq{k}, where $\partial = \frac{\partial }{ \partial z}=\frac 12(\p_x-\ii\p_y)$ and $\bar{\partial}
= \frac{\partial}{\partial \bar{z}}=\frac 12(\p_x+\ii\p_y)$ are the holomorphic and anti-holomorphic
derivatives, respectively. For a two dimensional diagonal metric, the  Ricci scalar curvature is  given by \eq{r}, where the Laplace-Beltrami operator is   \(\Delta_g=(4/\sqrt
g)\p\bar\p\). 
\begin{align}
&\text{Conformal gauge} & \quad & ds^2  = \sqrt g dz d \bar z \la{d}  \\
&\text{Ricci scalar curvature} & \quad  & R  = - \Delta_g \log \sqrt g \la{r}\\ &\text{K\"ahler potential} & \quad  &\partial \bar \partial K  = \sqrt{g}\la{k} \end{align}
 We will consider surfaces in the conformal class
of a sphere, which means that the metric can be written as \eq{gs}, where $\sqrt {g_0}$ is the metric on a uniform sphere  of radius $r$ with constant curvature $R_0$, whose  area 
\(4\pi r^2\) is equal to the area of the surface \(V\). The conformal factor  $e^{2 \sigma}$
describes deformations of the surface, and the function \(\sigma\) is regular on the extended complex plane (Riemann sphere).
\begin{align}
  \la{gs}  & \sqrt g  = e^{2 \sigma} \sqrt{g_0} &  \sqrt{g_0} & =  (1 + |z|^2/4 r^2)^{-2} \\
  \la{ks} &K  =  K_0 + 2u &  \quad K_0 & =\frac{V}{\pi} \log (1 + |z|^2/4 r^2) \\
\la{rs} &R  =  e^{-2\sigma}R_0  - 2e^{-2\sigma}\Delta_0\sigma & \quad   R_0 & =2 r^{-2} \end{align}
 Here, $u$ is a real-valued function which describes deformations of the K\"ahler potential $K_0$ of the uniform metric. It is related to the conformal factor $\sigma$ through $\Delta_{0} u = 2 (e^{2\sigma} - 1)$.  Also we  will use the mean curvature $\bar R = V^{-1} \int R dV = 4 \pi \chi / V$. For surfaces of constant curvature $\bar R= R_0$

The  Laplace-Beltrami operator  is \(\Delta_g=e^{-2\sigma}\Delta_0\), where \(\Delta_0\)  is the  the Laplace-Beltrami
operator on the sphere,  and the covariant derivatives are   $\nabla_z=\frac 1{\sqrt g}\p\sqrt g$ and $\nabla_{\bar z}=\frac 1{\sqrt g}\bar \p$ .

At infinity, the K\"ahler potential and metric functions behave as
\begin{align}
K& \to \frac{V}{\pi} \log |z|^{2}, \quad \log \sqrt{g} \to - 2 \log |z|^{2},\\
u &\to 0, \quad  \log \sigma \to 0.
\end{align}
This is due to our choice of coordinate charts. In our chosen coordinates, the point $z_{0} = \infty$ is a marked point, in the sense that this is precisely where our coordinate chart ``stops working". A M\"obius transformation would take this to any arbitrary point on the complex plane. 

Finally, it is useful to introduce the spin connection, which acts as a gauge field for local rotations. {In conformal gauge, it takes the form $\omega_{i} = \frac{1}{2}\epsilon_{ij} \partial_{j} \log \sqrt{g}$, or in complex coordinates $\omega = \frac{1}{2} (\omega_{1} - \ii \omega_{2}) =  (\ii/2) \partial \log \sqrt{g}$. The flux of the spin connection is just the curvature $\partial_{1} \omega_{2}- \partial_{2} \omega_{1} = \frac{1}{2} R \sqrt{g}$.

 \section{Quantum Hall States in a Curved Space}\la{Sec2}

We review the problem of free electrons in the lowest Landau level (LLL). 

\subsection{Integer Quantum Hall Effect on a Closed Surface }\la{IQHresults}

To model free electrons on a Riemann surface, we use the Pauli Hamiltonian for
spin polarized electrons with an arbitrary Land\'e ${\rm g}$-factor $g_{s}$
\begin{align}\la{H}
H = \frac{1}{2m}\left(\frac{1}{\sqrt{g}}\pi_a \sqrt{g} g^{ab} \pi_b- \frac{g_{s}}{2} e \hbar B\right),
\end{align}
where \(\pi_a=-i\hbar\p_a-eA_a\) is the kinetic momentum, and \(a,b=1,2\) \cite{Iengo1994, Maraner1992}.

In complex
 coordinates, the Hamiltonian reads
\begin{align}\la{Hholo}
H = \frac{2}{m}\left(\frac1{ \sqrt{g}} \pi\bar \pi+ \frac{2-g_{s}}{8}
e\hbar B\right),
\end{align}
where
\begin{align}\la{Pi}
\bar\pi=-\ii\hbar\bar\p-e\bar A. 
\end{align}
is the anti-holomorphic component of the momentum. Eq.\eq{Hholo} follows from the commutation relations    $[\bar \pi, \pi] =  e \hbar B \sqrt{g}/2$. In terms of the gauge potential $\mathbf A$, the magnetic field is  \(B \sqrt g = 2\ii (\bar{\partial} A - \partial \bar{A}) \).  We choose
the covariant Coulomb gauge  { \(\frac{1}{\sqrt{g}} \partial_{i} \sqrt{g} g^{ij} A_{i}  =\frac{1}{\sqrt{g}}(\bar{\partial} A + \partial \bar{A})=0\)}
and introduce a real potential \(Q
\) defined by   $ eA=\ii\hbar {\partial} {Q}/2$. The ``magnetic potential"
\(Q\) obeys the equation
 \begin{align}\la{Q} - \hbar\Delta_{g} {Q}= 2eB.\end{align} 
Then the kinetic momentum
\eq{Pi} reads\begin{align}
\bar\pi=e^{Q/2}(-\ii\hbar\bar\p) e^{-Q/2}.
\end{align}

\subsubsection*{Uniform Magnetic Field}First, we consider electrons on a curved surface in a  uniform
magnetic field. { For a uniform magnetic field, the field strength is proportional to the volume form, such that an infinitesimal magnetic flux is given by $BdV$ for constant $B$.} The states
in the LLL are zero modes of $\bar \pi$,
 \begin{align} \label{pi}
  \bar\pi \psi=0.  
  \end{align} 
In a uniform magnetic field, \(Q=-K/2l^2\), where
 \(l\) is the magnetic length and  \(K\) is the K\"ahler potential  (note
that   in flat space the  K\"ahler  potential is $K =  |z|^{2}$). The  solutions to \eq{pi} are the single particle
eigenstates 
\begin{align}\la{2.7}
\psi_n(\xi) = s_n(z) e^{-K(\xi)/4l^2},\quad \xi=(z,\bar z). 
\end{align}
where $\xi$ is a point on the surface. The notation is used to indicate that $K$ is not a holomorphic function of $z$. This is the convention we use throughout the paper. 

The  holomorphic functions  
\(\{s_n \}\) are defined as solutions to the equation
  \(\bar
\partial s_n = 0\) such that \(\psi_n\) is normalizable under the inner product { 
\begin{align}\la{2}
 \langle \psi_{n}| \psi_{m} \rangle \equiv \int s_{n} \bar{s}_{m}  e^{-K/2l^2}
dV = \delta_{mn},
\end{align}}
where $dV=\sqrt{g}d^{2}\xi = \sqrt{g} dx dy$. 
 In mathematical literature, $s_n$ are known as sections of the holomorphic line bundle 
 {equipped with a hermitian metric $e^{ - K/2l^{2}}$.

Holomorphic sections \(s_{n}\) defined in the conformal class of
a sphere are polynomials (have poles at infinity) subject to the condition \eq{2}. Bearing in mind the asymptotes \begin{align}
   K \to(V/\pi) 
 \log |z|^2  ,\quad \log\sqrt g\to -2\log |z|^2, \quad\text{at}\;z\to\infty, \la{2.71}
\end{align}   we find that the
degree of  \(s_n\) can not exceed \(N_\phi\),  the total flux in units of the flux quantum.  This is the Riemann-Roch theorem in a nut-shell. For an arbitrary genus the
arguments are similar. The number
of admissible states, which is
equal to the number of holomorphic sections on a manifold $M$ with genus $g$, is $N_1= N_\phi - g +1$.

The  many-body
 ground state wave function for free fermions is the Slater determinant of
the single particle eigenstates 
$$\Psi_1 ( \xi_1, ..., \xi_N) = e^{-\sum_i^N
\\ K(\xi_i)/4l^2} \det [s_n(z_i) ]/\sqrt{N!},$$
where the number of particles
must not exceed \(N_1\). 
 The wave function in this form
was studied in detail in Ref. \cite{Klevtsov2013}. 

If the many-body  state consists of    states with consecutive degrees \(n=0,1,\dots,N\),
the
 Vandermonde identity implies
\(\det
[s_n(z_i) ]  =\sqrt{N!} \mathcal{Z}^{-1/2}\Delta \), where \begin{align}\label{Delta}
\Delta=  \prod_{i < j}^N
(z_i - z_j),
\end{align} 
and the wave function takes the form \cite{Klevtsov2013}
\begin{align}\label{wf2}
\Psi =\mathcal{Z}[g]^{-1/2}\Delta\, e^{ - 
\sum_{i} K(\xi_i)/4l^2},
\end{align}
which is valid for the spherical, conical,  and planar geometries. Generally,  $\det[s_{n}(z_{i})]$ is determined entirely by the topology of the manifold. For example,  for toroidal geometries, periodicity requirements constrain $\det[s_{n}(z_{i})]$ to be written in terms of elliptic theta functions. However, the short distance behavior as $z_i \rightarrow z_j$ must always be of the form \eq{Delta}.

In the
  case that $N <N_1$, the electronic droplet continuously
 occupies
the surface leaving an unoccupied area centered around the marked point \(z_0=\infty\). If the number of particles in the state is $N_{1}$, then the droplet occupies the entire surface and the Landau level is
completely filled.  

The normalization factor \(\mathcal{Z}[g]\)  is a functional of the metric and the number of particles. We will see that this functional (which we call a generating functional) is an essential characteristic of QH states.

\subsubsection*{Non-Uniform Magnetic Field} In a non-uniform
magnetic
field and general $g_{s}$, the degeneracy of the lowest
Landau level breaks and, generally, the wave function \eq{wf2} is no longer accurate.  However, for $g_{s} = 2$, the LLL remains degenerate, which is evident from  \eq{Hholo} \cite{AC79}, and the eigenstates are once again zero modes
of  $\overline{\pi}$
 \eq{pi}. The single particle eigenstates are $\psi_n(\xi) = s_{n}(z)e^{   {\frac 12Q(\xi)}}
$, where  $Q$ is defined by \eq{Q}. { As $z \rightarrow \infty$, $ Q\sim- N_\phi \log |z|^2$, and the number of admissible
states  is  \(N_\phi+1\) as before}. The many-particle
state constructed now reads
\begin{align}\la{wf_mag}
\Psi=\mathcal{Z}^{-1/2}\,\Delta\, e^{   \frac 12\sum_{i}
Q(\xi_i)}.
\end{align}
 In an experimental setting, the $g-$factor often differs from $2$ and the Landau
level splits. However,  if the magnetic field varies slowly over a large constant background, the wave function \eq{wf_mag} is still a valid approximation for modeling geometric effects.

\subsubsection*{Spin}  A possible
 generalization of  states on  the lowest Landau level includes the spin. In this case the kinetic momentum \eq{Pi} is  modified by  the spin connection 
\begin{align}
  \bar{\pi}=-\ii \hbar \bar{\partial}  + \hbar s \bar{\omega}-  e\bar{A},\quad  \bar{\omega}  =
-(\ii/2) \bar{\partial} \log \sqrt{g}
\end{align}

 The spin $s$ modifies the inner product by changing $dV\to (\sqrt g)^{-s}dV$ in \eq{2}
The holomorphic sections remain polynomials  on a surface of
genus $0$, but the degeneracy of the lowest Landau level changes to  $N_{\phi} + 1 -2 s $.  

The wave function with spin will pick up an adiabatic phase $(s/2) \int R dV$ upon traversing a closed path on a curved surface. The integral measures the total rotation of a local frame after parallel transporting along the contour. Thus, a $2\pi$ rotation leads to a phase change $2\pi s$.

 Apart from  being an additional  useful parameter, spin may be relevant to certain QH states. For example, spin $s=1/2$ corresponds to the systems with particle-hole symmetry such as graphene or possibly to other fractions beside Laughlin's series.

\subsection{Fractional Quantum Hall Effect on a Closed Surface}
 The QH wave function  \eq{wf_mag}
is the product of two distinct parts {- the real-valued exponential factor \(\exp\left(  \frac 12\sum_{i} {Q}(\xi_
i)\right)
\) (which becomes the Gaussian factor in flat space) and a holomorphic polynomial $\prod_{i<j}(z_{i} - z_{j})$. The exponential factor depends  on the magnetic field and the metric but not on the QH state.  If the magnetic field is uniform, $Q$
  is the K\"ahler potential and the exponential factor is determined entirely by the underlying geometry.}

The remaining part of the wave
function is  an antisymmetric
 (fermions) or  symmetric (bosons) homogeneous holomorphic function 
which vanishes when particles coincide. The wave function must possess this general structure  for the FQH case restricted to the lowest Landau level, which constrains it to the form
\begin{align}\la{LLL_wf}
\Psi = \frac{1}{\sqrt{\mathcal{Z}[g]}}F(z_{1}, .., z_{N}) \exp\left(  \frac 12\sum_{i} {Q}(\xi_
i)\right).
\end{align}
The holomorphic function  $F(\{ z_{i}\})$ distinguishes different QH states through the filling fraction and shift. We consider only the case of genus-0 surfaces, for which $F$ is a homogeneous holomorphic
polynomial. This follows by requiring the probability density transform co-variantly under M\"obius transformations, discussed below. We assume that $F$ is normalized such that the coefficient in front of its  highest monomial is 1 as in \eq{wf_mag}. This uniquely   defines the overall normalization factor $\mathcal{Z}[g]$.

The probability
density
of the QH state is
\begin{align}\la{13}
dP=|\Psi|^2\prod_i dV_i = \frac{1}{\mathcal{Z}[g]}|F|^2\prod_{i=1}^Ne^{W(\xi_i)}d^2\xi_i
\end{align}
where  \begin{align}\la{141}
W=\frac 12 (Q+\bar Q)+(1-s)\log\sqrt g.
\end{align}

When the magnetic field is uniform, this reads  
\begin{align}\la{2.17}W = -K/2l^2 + (1-s)\log \sqrt{g}.\end{align}

 The potential \(W\) can be written with reference to a constant curvature surface (sphere) using the relation $N_{\phi} = V/2\pi l^{2}$, 
 \begin{align}W=&w[\sigma]- (N_{\phi} - 2s + 2)\log (1+|z|^2/4r^{2}),\\ &
 \la{66}w[\sigma]= -u/l^2 +2(1-s)\sigma,\end{align}
 
where    \(2u=K-K_0\)
 is the difference between the  K\"ahler potential
 of the  deformed sphere  and the uniform sphere , and \(\sigma\) is the conformal factor \(\sqrt
g=e^{2\sigma}\sqrt{g_0}\).

The probability
density
 is then

\begin{align}\label{14}
dP= \frac{1}{\mathcal{Z}} |F|^2
\prod_{i=1}^N e^{ w(\xi_i) }\left(1+|z_i|^2/4r^{2}\right)^{-\left(N_{\phi} - 2s + 2\right)}d^2\xi_i
\end{align}

  All of the geometric information regarding the deviation  from the sphere is then contained in the \( w[\sigma]\) term.

\subsubsection*{Coherent states and geometry}The formulas  for the probability density \eq{13} and \eq{14} admit the following interpretation: the
ground state on a  deformed sphere
  is equal to a coherent  state on a  round
sphere.  

We clarify this correspondence on the most  familiar case of a plane. A coherent state is  constructed by acting with the normally ordered operator \(e^{\frac 12\sum_iw(2l^2\p_{z_i},z_i)}\) on
the holomorphic part of the ground state \(e^{\frac 12\sum_iw(\p_{z_i},z_i)}F\). Under the Bergmann  inner product  \(\langle
F|F'\rangle=\int e^{-|z|^2}\overline {F(z)} F'(z) d^2\xi\)  this operator acts as a multiplication by  the factor $e^{\frac 12\sum_i w(\bar z_i,z_i)}$(see
e.g. \cite{girvin1986magneto}) due to the identity \( e^{ |z|^2}\p e^{-|z|^2}=
-\bar z\).
Therefore, a modification of the weight   \(e^{-|z|^2/2 l^2}\to e^{-|z|^2/2l^2}e^{w(z,\bar z)}\) in \eq{13} could be understood either as a coherent state in a flat
geometry or a ground state  on a surface with the metric found from  \eq{66}. 

  \subsubsection*{M\"obius Invariance  and the Maximum Number of Particles}\la{221}
The form of the holomorphic polynomial $F$ for FQH states is significantly constrained by M\"obius invariance. In the limit of large $N$, the mean particle density forms a droplet with a finite support. In the planar geometry \cite{Zabrodin2006}, the density forms a droplet with a boundary. However, on a compact surface, it is possible for the droplet support to be the entire surface. This occurs for a particular value of $N$, as determined by M\"obius invariance below. 

For a droplet which covers the surface, there is no preferred marked point. However, the K\"ahler potential  depends explicitly on the
choice of the point  (set to infinity) specified by  \eq{2.71}. This point is moved by
the M\"obius transformation, under which    the K\"ahler potential  and the metric transform as 
\begin{align}
 z\to f(z) = \frac{az+b}{cz+d},\quad \sqrt g\to\left|f'(z)\right|^2\sqrt
g,\quad  K\to K+
\frac{V}{\pi}\log\left|f'(z)\right|.\la{trans_mob}
\end{align}  
 As the result of the transformation, the real exponential term in \eq{13} \(e^Wd^2\xi\)
acquires the factor \(\left|f'(z)\right|^{-({N_\phi}-2s)}\). This factor  must be compensated by the transformation of the holomorphic polynomial
\(F\)  in \eq{LLL_wf}. Therefore the function \(F \) must be primary under the M\"obius transformation
 \begin{align}\la{23} 
F(\{f(z_{i})\}) & = \prod_{i = 1}^{N} \big(f'(z_{i})\big)^{-h_{N}} F(\{z_{i}\})
 \end{align} 
with a holomorphic  dimension given by  
 \begin{align}\la{3}
-2h_N= N_\phi-2s. 
\end{align}

This property links the holomorphic functions \(F\) to polynomial
correlation functions  of primary operators of Conformal Field Theory. 

The dimension \(h_N\) depends on the total number of particles and characterizes the QH state. 
For $-2 h_{N} < N_{\phi}-2s$, the M\"obius transformation shows a vanishing of the probability density near the marked point at infinity. This indicates the presence of a puncture, or boundary, in the particle density. For $-2 h_{N} > N_{\phi}-2s$, the wave function is not normalizable. 


\subsubsection*{Filling fraction and the shift}
The pair-wise interaction between particles suggests that the  dimension \(h_N\)  rises linearly
with  the number of
particles. We denote it as
 \begin{align}\la{2.23}
-2h_N=\nu ^{-1}N-\mathcal{S},
\end{align} 
The formula defines  two \(N\)-independent coefficients \(\nu\) and  \(\mathcal{S}\)
 which characterize the state. They
 are  the filling fraction and the shift, respectively. The constraint Eq. \eq{3} imposed by M\"obius invariance of the probability density, combined with the dimension \eq{3} leads to the condition \eq{shift-spin} at \(\chi=2 \) and zero spin $s$. When modified by spin, we find the maximum number of particles which can be placed on a compact surface is given by \eq{shift-spin}.

\subsection{FQH Model Wave Functions}\la{fqh_model_wf}
An important characteristic  of  the  wave function is given by the positions and the  degree of the zeros as two particles coincide. For the Laughlin state \cite{Laughlin1983}, the zeroes of the wave function are located at the particle coordinates, and the wave function vanishes as $(z_{i} - z_{j})^{\nu^{-1}}$ as any two coordinates merge. In this case, the degree of the zeroes is simply related to the filling fraction. In  other  FQH states this is not necessarily the case. In general, we can have \(
 F(z_{1}, z_{2}, ..., z_{N}) |_{z_1\to z_2}\sim (z_{1} - z_{2})^{m},
\)
where $m \le \nu^{-1}$. This means that there  exist  configurations which cause the wave function to vanish, but in which no two particles coincide. This is the case, for example, for the Pfaffian states \cite{Moore1991} and more general  parafermion states  \cite{Read99} summarized in \ref{A}.

\subsubsection*{Laughlin state}
In the Laughlin state the inverse filling fraction, the shift and the degree of zeros are the same. We denote it as $\beta$
\begin{align}\text{Laughlin states}\quad m=\mathcal{S} =\nu^{-1}\equiv \beta.
\end{align}
This data fixes 
\begin{align*}
-2h_{N} = \beta (N - 1),
\end{align*}
according to Eq.\eq{2.23}, and uniquely specify the holomorphic piece (up to a constant) to be
\begin{align}\la{liouville_F}
F(z_{1}, ..., z_{N}) = \prod_{i<j} (z_{i} - z_{j})^{\beta}.
\end{align}

\subsubsection*{Pfaffian states}
The   Pfaffian state \cite{Moore1991} takes the form
\begin{align}\la{pfaffian_wf}
F = {\rm Pf}\left( \frac{1}{z_{i} - z_{j}}\right)  \prod_{i<j}(z_{i} - z_{j})^{M+ 1}, 
\end{align}
where \({\rm Pf}(M_{ij})\) is the Pfaffian of an anti-symmetric matrix $M_{ij} = 1/(z_{i} - z_{j})$. For odd $N$ the Pfaffian vanishes. The filling fraction for these states is $\nu = 1/ (M+1)$. The dimension of the Pfaffian is $ -1/2$, and the total dimension of the state is 
\begin{align*}
 -2h_N=(M+1)(N-1)-1.
 \end{align*}
This constrains the shift and the filling fraction to the values
\begin{align}
\text{Pfaffian
states}\; &&\nu^{-1} = {M + 1}, &&\mathcal{S}=\nu^{-1} + 1 = M+2.
\end{align}

\section{Generating Functional and Linear Response Theory }\la{Sec3}
Using the general form of LLL wave functions \eq{LLL_wf}, we can make some
general statements about near-equilibrium response. The basic tool, which
is one of the central objects of study in this paper, is the {\it generating
functional}
\begin{align}\label{gen_fun}
\mathcal Z[W] = \int  |F(z_{1}, ..., z_{N})|^{2} \prod_i^N e^{ W(\xi_i)}d^2\xi_i,
\end{align}
where a polynomial \(F\) depends on the QH state, but does not depend on
magnetic field and the curvature of the surface. We normalize the polynomial
such that the the coefficient in
front of the leading monomial at infinity is 1.

{ The expectation value  of any operator 
\(\mathcal{O}(\{\xi_i\}) \), where $\mathcal{O}$ is symmetric, is}
\begin{align}\la{32}
\langle \mathcal{O}\rangle=\int  \mathcal{O} dP=\mathcal{Z}^{-1}[W]  \int  \mathcal{O\,}|F|^{2}
\prod_i^N e^{ W(\xi_i)}d^{2}\xi_i,
\end{align} 

  The   generating functional $\mathcal Z [W]$ defined in \eq{gen_fun} is useful for studying transport coefficients, because the variations thereof over $W$ yield correlation functions of the density,
  \begin{align}\la{3.4}
\langle  \rho (\xi)\rangle
= \frac 1 {\sqrt {g(\xi)} 
} \frac{\delta \log \mathcal{Z}}{\delta W(\xi)},  \qquad   \langle  \rho(\xi) \rho (\xi')\rangle_{c} = \frac 1 {\sqrt {g(\xi)} \sqrt {g(\xi')}}\frac{\delta^{2} \log \mathcal{Z}}{\delta W(\xi) \delta
W(\xi')}   .
 \end{align}
where $\rho(\xi) = \sum_{i = 1}^{N} \frac{1}{\sqrt{g}}\delta^{(2)}(\xi- \xi_{i})$. Here, the connected correlation function of two operators is defined $\langle
A B \rangle_{c} \equiv \langle A B\rangle - \langle A \rangle \langle B \rangle$, 
 and variations over $W$  do not change the conformal class and are equivalent to Weyl rescaling, since $\delta W=\delta w$ as in \eq{66}.  

 Similarly \begin{align}\la{rhoW}
\frac{\delta \langle \mathcal{O}\rangle}{\delta W(\xi)}= \sqrt{g} \langle
\mathcal{O\,}  \rho(\xi) \rangle.
\end{align}
 This method for computing correlation functions  was employed
in \cite{Zabrodin2006}.
 
  Since the the function \(W\) is linked to geometry  by Eq. \eq{2.17}, we observe that the
 correlation functions can be interpreted as a response to the  variation
 of the geometry.
 
\subsection{Dilatation and Translation Ward Identities (Sum Rules)}\la{2.4} The transformation law \eq{23} is a source of two  sum rules. They are the Ward identities with respect to the
sub-algebras of the M\"obius transformations - translations and dilatations.

 Consider the expectation value  of an operator $\mathcal O$.
 In  \eq{32} we  change the integration variables by an infinitesimal M\"obius  transformation: translation \(z_i\to
z_i+\epsilon\),  dilatation transformation \(z_i\to z_i+\epsilon z_i\), and special conformal transformation  \(z_i\to z_i+\epsilon z_i^2\), then compute the change of the integrand to leading order \(\epsilon\).  Obviously, the value of the integral does
not change. Therefore the integral of the transformation  of the integrand vanishes.
   According \eq{23} we obtain 
   \begin{align}\nn&  2(1-h_N)\sum_i\langle \mathcal{O}z_i\rangle - \langle
\mathcal{O}\ell_{+1}W\rangle= \langle\ell_{+1}\mathcal{O}\rangle,&\ell_{+1}=-\sum_i
z_i^2\p_{z_i},
\\
\la{2.29} & N(1-h_N)\langle \mathcal{O}\rangle - \langle
\mathcal{O}\ell_0W\rangle= \langle\ell_0\mathcal{O}\rangle,&\ell_0=-\sum_i
z_i\p_{z_i},\\
\nn
& -\langle \mathcal{O}\ell_{-1}W\rangle= \langle \ell_{-1}\mathcal{O}\rangle, & \ell_{-1}=-\sum_i
\p_{z_i}, \end{align} 
where  \(\ell_{+1},\,\ell_0\) and \(\ell_{-1}\) are the generating operators of special conformal transformations, dilatations and
translations. Here, $\ell_{k}W \equiv \ell_{k} \sum_{i} W(\xi_{i})$. 

 In particular, setting \(\mathcal{O}\) to 1 and passing to the limit of large number of particles where we replace the sum by the
integral with the density we obtain the sum rules \begin{align}\la{3.9} 0= \int
(z^2\p_z W+2(1-h_N) z)\langle\rho\rangle dV,\quad -N(1-h_N)= \int
z\p_z W\langle\rho\rangle dV,\quad 0= \int
\p_z W\langle\rho\rangle dV.\end{align}
In \ref{V} we check that the sum rules are saturated by the formula
\eq{1}.

\section{Response to Variation of Curvature } \la{Sec4}
 We define the {\it density response to curvature} 
\begin{align}\la{response_R}
\eta(\xi ,\xi') &\equiv  \frac{2\pi}{\nu} \frac{\delta \langle
\rho(\xi) \rangle}{\delta R(\xi')}. \qquad  
\end{align}
 Evaluated  in the flat space at \(R=0\), the response function
is translation invariant \(\eta(\xi ,\xi')=\eta(\xi - \xi')\). Then we can
pass to Fourier space \begin{align}
\eta(k) =  \int e^{ -\ii {\bf k} \cdot \xi  } \eta(\xi )d^{2}
\xi.
\end{align} 
It follows from \eq{1} and  \eq{shift-spin} that the zero momentum value of \(\eta\) is given in terms of spin and shift by 
\begin{align}\la{eta}
\eta(0)=\frac 14(\mathcal{S}-2s).
\end{align}

The response of the density to the curvature is intrinsically related to the
odd viscosity, introduced in \cite{Seiler1995}. We reserve this discussion for \ref{B1}. Here, we state the result, which relates the homogenous parts of the anomalous viscosity $\eta^{(A)}$ to the density response to curvature
\begin{align}\eta^{(A)} = \hbar \eta (0).
\end{align}

\subsection{Response to Variation of Curvature and the Structure Factor}\la{hallviscosity_structurefactor}

Let the magnetic field be uniform. Then $W= -K/2l^{2} +( 1 - s) \log \sqrt{g}$. A variation of $W$ is then equivalent to a variation of the metric or equivalently variation of the  K\"ahler
potential  \(-2l^{2}\delta W=(1 -\frac{1}{2}(1-s)l^{2}\Delta_g)\delta
K\).   Specifically, 
\begin{align}\nn
\delta \langle \mathcal{O} \rangle &= \int dV\langle \mathcal{O\,} \rho(\xi)\rangle_{c} \,  \delta W(\xi)\\
&  =-\frac 1{2l^{2}} \int dV \,  \delta K(\xi) \left( 1 - \frac{1}{2}\left( 1 - s\right) l^{2}\Delta_{g}\right) \langle \mathcal{O}\rho(\xi)\rangle_{c}
\end{align}
Assuming that the variation of the metric preserves the area, we find that
\begin{align}
- 2 l^{2}\frac{1}{\sqrt{g(\xi)}}\frac{\delta \langle \mathcal{O}\rangle}{\delta K(\xi)} = -\frac{1}{2}l^{2}\Delta_g\frac{\delta\langle \mathcal{O}\rangle}{\delta \sqrt{ g(\xi)}} =\Big  ( 1 - \frac{1}{2}\left(1-s\right) l^{2}\Delta_g\Big)\langle  \mathcal{O\,} \rho(\xi)\rangle_c. 
\end{align} 
While this is generally true, it has particularly interesting consequences for the choice of \(\mathcal{O}= \mathcal{Z}\) and $\mathcal{O} =\sqrt g\rho$. With the former, it reads
\begin{align}\la{3.7}
& -2l^{2}\frac{1}{\sqrt{g}}\frac{\delta \log\mathcal{Z}}{\delta
K} =\Big ( 1 - \frac{1}{2}\left( 1- s\right) l^{2}\Delta_g\Big)\langle 
\rho\rangle.
\end{align}

This relation connects geometric variations of the generating functional to the particle density. In Sec.\ref{Sec9}, we start with the density $\langle \rho\rangle$ to determine the asymptotic expansion of generating functional. 

For the choice of $\mathcal{O} = \sqrt{g} \rho$, we have
\begin{align}
&-\frac{l^{2}}{2} \Delta_{g, \xi'} \left(\frac { \delta (\sqrt{g} \langle
\rho(\xi)\rangle)}{ \delta \sqrt{g(\xi')}}  \right) \!= \!\left( 1- \frac{1}{2}\left( 1 - s\right)l^{2}\Delta_{g, \xi'}\right) \langle \rho(\xi) \rho(\xi') \rangle_{c} \la{var_rho_1}
\end{align}
This formula connects the two-point density function to the variation of
the density over  the metric. It is particularly meaningful to consider variations away from the flat
metric $\sqrt{g} \approx 1 + 2\sigma$, with curvature $  R \approx
-2 \Delta\sigma$, where $\Delta = 4\partial \bar{\partial}$ is the Laplacian. Then to leading order in $\sigma$, \eq{var_rho_1} can be written as
\begin{align}\la{var_rho_2}
- \frac{1}{2}l^{2} \Delta_{\xi'} \left( \delta^{(2)}(\xi \!-\! \xi') \langle \rho(\xi) \rangle -\Delta_{\xi'} \frac{\delta \langle \rho(\xi) \rangle}{ \delta R(\xi')}  \right)\! =\! \left( 1\!-\!\frac{1}{2}\left( 1- s\right)l^{2}\Delta_{ \xi'}\right) \langle \rho(\xi) \rho(\xi') \rangle_{c}.
\end{align}
 
  Let us  recall the definition of the {\it static structure factor}.  In the flat space, the connected two-point density correlation function is translation invariant.  The static structure factor is then given by the Fourier transform
   \begin{align}\la{3.11}
S(k) = \frac{V}{N} \int e^{ - i {\bf k} \cdot {\bf \xi }} \langle \rho(\xi) \rho(0)\rangle_{c}d^2\xi.
\end{align}
  Eq \eq{var_rho_2} results  in a relation between the density response to curvature and the structure
factor. Writing the momentum $k$ in units of inverse magnetic length $l^{-1}$, it reads
 \begin{align}\la{eta_sk}                                       
\frac{k^{2}}{2} + \frac{k^{4}}{2} \eta(k) = \left(1 + \frac{k^{2}}{2}\left( 1 - s\right)\right) S(k).
\end{align}
This connection generally holds for states in the lowest Landau level.
We emphasize that the relation  does not depend on the choice of the  QH state.

An immediate application  of this  relation and the formula \eq{eta} is the value of  the  first two terms of the
gradient expansion of the structure factor 
\begin{align}\la{s}
\!\!S(k)= \frac {k^2}2\! +\!\left(\mathcal{S}-2\right)\frac {k^4}8+\dots
\end{align}    
The leading term in the structure factor does not depend on the QH state. This is a consequence of conservation of particle number and angular momentum \cite{girvin1986magneto,girvin1985collective}. The  next to the leading coefficient is related to the anomalous  viscosity  and 
vanishes for   QH states
  with  \(\mathcal{S}=2 \) \cite{Read2011}. 
  
We comment that the projected static structure factor, defined in \cite{girvin1986magneto} as $\bar{S}(k) = S(k) - 1 + \exp(- k^{2}/2)$, has the leading behavior
\begin{align}\la{projected_s}
\bar{S}(k) = \frac{\mathcal{S} - 1}{8} k^{4} + ...,
\end{align}
saturating the lower bound for the leading coefficient proposed in \cite{Haldane2009, Golkar2013}. This follows directly from the form of the LLL wave function \eq{LLL_wf}.
 
The density correlation function  evaluated in flat space is a scalar with respect to a rotation and for that reason does not depend on  spin. Thus the formula  \eq{eta_sk}  explicitly shows the spin  dependence of $\eta(k)$.









\subsection{The Hall Conductance, the Structure Factor and a Response to
a Variation of Magnetic Field}\la{hallcond}
Let us now assume a flat space but consider  a weakly inhomogeneous magnetic field \(B=B_0+\delta B\), on the background of a large  uniform field. According
to the St\v reda formula \eq{0}, the response of the density to a variation of magnetic
field is the Hall conductance
\begin{align}\la{response_B}
\sigma_{H}(\xi -\xi') \equiv \frac{e\delta \langle \rho(\xi)\rangle}{\delta B(\xi')} \Big|_{B = B_{0}}, \qquad
\sigma_{H}(k) = \int e^{- i {\bf k} \cdot \xi} \sigma_{H}(\xi)d^{2} \xi,
\end{align}
According to \eq{Q} a variation of the magnetic field is described by the
variation of the function \(Q\) as \(\hbar\Delta\delta Q=-2e\delta B\).
 Differentiating the generating functional as in \eq{3.4} we obtain a relation between the Hall conductance
and the structure factor
\begin{align}
\Delta_{\xi'} \frac{\delta \langle \rho(\xi)\rangle}{\delta B(\xi')} = - \frac{2e}{\hbar} \frac{\delta \langle \rho(\xi)\rangle}{\delta W(\xi')} = - \frac{2e}{\hbar} \langle \rho(\xi )\rho(\xi')\rangle_{c},
\end{align}
and in Fourier space,
\begin{align}\la{3.15}
 \sigma_{H}(k) = \frac{\nu e^{2}}{h} \frac{2 S(k)}{k^{2}l^{2}}.
\end{align}
We found the earliest appearance of the relation between the structure factor and the Hall conductance at $k = 0$ in \cite{McMullen1985}. The formulas \eq{eta_sk}   and \eq{s} imply that the uniform part of the  Hall conductance is  quantized in units of
the  filling factor \(\sigma_{H}(0) = {\nu e^{2}}/{h}\), and also determine the next to the leading correction 
\begin{align}\nn
\frac{\sigma_H(k)}{\sigma_H(0)}=\!1\!+\! \frac{(\mathcal{S} - 2)}{4} k^{2} + ...
 \end{align}
This correction was found in \cite{Bradlyn2012,Hoyos2012}, and shown to be related to the anomalous viscosity. We emphasize that the relation \eq{3.15}  holds only for  $g_{s} = 2$, when the LLL remains degenerate at a non-uniform magnetic field. The Hall conductance
receives non-universal corrections  already of the order $k^{2}$  in realistic devices, where the Land\'e g-factor is known to deviate  from $2$.
These
corrections reflect a contribution of  diamagnetic currents.   However, these corrections are suppressed by the ratio
between the gap in one-particle excitations and the cyclotron energy. Note that we can also write \eq{3.15} in terms of the spin dependence of the curvature response as $\sigma_{H}(k)/\sigma_{H}(0) =- 2\partial_{s} \eta(k)$.
In the curved space the St\v reda formula \eq{response_B} and 
\\
\\
In summary, we see that the transport coefficients of the FQH states are determined
by the response of the system to the variation of the spatial geometry.
\section{Ward Identity }\la{Sec6}
In this section, we present the techniques for computing  expectation
values and correlation functions of  FQH 
states. 

Our primary tool of inquiry is a Ward identity, also called the loop equation
in  Random Matrix Theory literature. The method was developed in detail in Ref. \cite{Zabrodin2006}. 

The invariance of the generating functional under holomorphic diffeomorphisms yields a Ward identity that yields an exact formula relating one-point and two-point correlations functions. The {sum rules generated by M\"obius transformations in Sec.\eq{2.4}}
 comprise parts of the Ward Identity. In fact, an infinite sequence of exact relations can be generated which connect $n$ and $n+1  $ -point correlation functions. Typically,  such sequences are truncated by some physically motivated assumption. In our case,  the  sequence is truncated by the short distance regularization of the two-point function.

\subsection{Stress Tensor}  

The stress tensor is an essential object for deriving  the Ward identity. The holomorphic  component of the stress tensor is defined as an operator  {  
 \begin{align}\la{5.11}
\widehat{\mathcal{T}}(z)=-\sum_i\p_{z_i}  \frac
1{z-z_i} ,
\end{align}  
}
    generating  holomorphic diffeomorphisms. 
Now consider the FQH states' generating functional 
\begin{align}\nn
\mathcal{Z}[g] = \int |F(z_1,\dots,z_N)|^2\prod_{i= 1}^{N} e^{W(\xi_i)}d^{2}\xi_i,
\end{align}
and perform a holomorphic re-parameterization of coordinates $z_i \rightarrow z_i + \epsilon / (z - z_{i}) $. The integral must be invariant under coordinate re-parameterization, { which is the source of the Ward identity.}  Keeping only the leading terms in the $\epsilon$ expansion, and taking the difference  between the original and re-parameterized generating functional yields {  
\begin{align}
\int \widehat{\mathcal{T}}(z)dP=- \int \prod_{k} d^{2}\xi_k\sum_{i} \frac{\partial}{\partial z_{i}} \left(\frac{1}{z - z_{i}}|F|^2e^{\sum_iW(\xi_i)}\right)=\int  dP\,\mathcal{T}_{-}(z) =0,
\end{align}
}
where  {  
\begin{align}\la{set_projection}
\mathcal{T}_{-}(z) =-|\Psi|^{-2} \sum_{i} \nabla_{z_i} \frac{1}{z - z_{i}} |\Psi|^2=-\frac{\p_{z_i}
W}{z-z_i}-\frac{1}{z-z_i}\p_{z_i}\log F-\frac{1}{(z-z_i)^2}
\end{align}
}
is a projection of the stress tensor onto the ground state and $\nabla_{z} = \frac{1}{\sqrt{g}}\partial \sqrt{g}$. This notation reflects the fact that the positive modes of $\mathcal T$ are annihilated when projecting onto the ground state.

Thus  the    Ward identity
reads\begin{align}\la{5.18}\langle \mathcal{T}_{-}(z)\rangle=0,\end{align}
or equivalently
\begin{align}
\Big\langle\sum_i\left[\frac{\p_{z_i}
W}{z-z_i}+\frac{1}{z-z_i}\p_{z_i}\log F+\frac{1}{(z-z_i)^2}\right]\Big\rangle=0.\la{5.1}
\end{align}

 \subsection{Ward Identity for Laughlin States}
 From this point on, we focus on the Laughlin states. In this case 
\begin{align}\nn
& \sum_{i}\frac{1}{z-z_i}\p_{z_i}\log F=\sum_{j\neq i}\frac{\beta}{(z-z_i)(z_i-z_j)}. 
\end{align}
After using the identity
\begin{align}\nn
\sum_{i \ne j} \frac{2}{(z - z_{i})(z_{i} - z_{j})} = \sum_{i, j} \frac{1}{(z - z_{i})(z - z_{j})} - \sum_{i} \frac{1}{(z - z_{i})^{2}},
\end{align}
we obtain\begin{align}\nn
\sum_{ i}\frac{1}{z-z_i}\p_{z_i}\log F=\frac{\beta}{2}
\left( \sum_{i} \frac{1}{z - z_{i}}\right)^{2} - \frac{ \beta}{2} \sum_{i}
\frac{1}{(z - z_{i})^{2}}.
\end{align} 
Applying these identities to \eq{5.1}, we obtain {  
\begin{align}\la{5.5}
- \mathcal{T}_-(z)= \sum_{i} \frac{\partial_{i} W(\xi_{i})}{z - z_{i}} + \frac{\beta}{2} \left( \sum_{i} \frac{1}{z - z_{i}}\right)^{2} + \frac{(2 - \beta)}{2} \sum_{i} \frac{1}{(z - z_{i})^{2}}.
\end{align}
}
Now we define  the field  $ \varphi$  { 
 \begin{align}\la{4.4}
  \varphi 
= - \beta \sum_{i} \left ( \log\left|\frac{z - z_{i}}{\sqrt{V/\pi}}\right|^{2} - \frac { \pi} V K(z_i) \right ).
\end{align}
 Physically, this field is the density potential, satisfying the Poisson equation $- \Delta_{g} \varphi = 4\pi \beta \rho$. The additional term proportional to the K\"ahler potential enforces the condition $\int \varphi(z) dV = 0$ to leading order. This definition is used for later convenience.}
 We write the Ward identity in terms of this field. Then Eq. \eq{5.5} reads {  
\begin{align}
\label{loopint}
 -\mathcal{T}_-(z)=\int \frac{  \partial W(\xi')}{z - z'}  \rho(\xi')  
 dV'  +  \frac{1}{2\beta}(\partial \varphi )^2  + \frac{2 - \beta}{2\beta}
  \partial^2 \varphi .
\end{align} 
}
This form  was obtained in \cite{Zabrodin2006} for general \(W\).

The field $\varphi$ is a generating function of collective coordinates
(or modes) \begin{align}a_{-k}=\sum_i z_i^{k},\la{a}\end{align} as 
{  \begin{align}\la{4.6}
\partial \varphi=- \frac{\beta N}{z} - \frac{\beta}{z} \sum_{k>0}a_{-k} z^{-k}.
\end{align} 
Similarly, $\bar{\partial}\varphi$ generates the anti-holomorphic modes $\bar{a}_{-k}$.}

The  projected stress tensor  
{  \begin{align}
\mathcal{T}_-=-\sum_{k\geq -1}z^{-k-2}\mathcal{L}_k,
\end{align}
}
generates the operators
\begin{align}\mathcal{L}_k=-\ell_{k}W +\frac\beta 2\sum_{n\geq
0}^ka_{-n}a_{n-k}+\frac{2 - \beta}{2}(k+1)a_{-k},\quad k\geq -1.
\end{align}
 where $\ell_{k}  \equiv - \sum_{i} z_{i}^{k+1}\partial_{i}$ are generators  of the Witt algebra.  Here $\ell_{k} W \equiv  \ell_{k} \sum_{i} W(\xi_{i}) $. Note that for $k = -1$, only the first term is non-vanishing. 
Then the Ward identity is a set of equations sometimes referred as Virasoro constraints \begin{align}
\mathcal{\langle L}_k\rangle=0 \quad k\geq -1.
\end{align}
 The first two of these identities are the sum rules (\ref{2.29}). Taking the expectation value of \eq{loopint}, 
 the Ward identity \eq{5.18} for the Laughlin state is  {  
 \begin{align}\la{5.8}
-\langle\mathcal{T}_{-}\rangle=\int \frac{  \partial W(\xi')}{z - z'}  \langle\rho(\xi')\rangle  
 dV'  +\frac{1}{2\beta}\langle\partial \varphi   \rangle^2+ \frac{2 - \beta}{2\beta}
  \partial^2\langle\varphi\rangle+  \frac{1}{2\beta}\langle(\partial
\varphi )^2  \rangle_c=0,
\end{align} }
where \(\langle ( \partial \varphi)^2 \rangle_{c}=\langle ( \partial \varphi)^2
\rangle-\langle ( \partial \varphi)\rangle^2\)  is the connected part of
the correlation function of two operators \(\p\varphi \)  at merged
points. 

 \subsubsection*{Anomalous part of the stress tensor}
The holomorphic component of the stress tensor $\mathcal {T}_{-}$ is not analytic. { Its definition \eq{5.11} implies that 
 $\mathcal T_{-}$ has poles at the positions of particles.}
Taking the covariant anti-holomorphic derivative $\nabla_{\bar{z}} = \frac{1}{\sqrt{g}}\bar\p$ from \eq{loopint}, we single out the non-analytic part 
{  
\begin{align}\la{5.81}
\frac 1\pi\nabla_{\bar{z}} \mathcal{T}_{-}=\rho (\partial   \varphi - \partial
W)   +   \frac{1}{2}(2 - \beta)
\nabla_{z}  \rho, \quad \nabla_{z} =\frac{1}{\sqrt{g}} \partial \sqrt{g}.
\end{align}
}
 Since the  expectation
values of \(\mathcal{T}_{-}\) and \(\bar \p \mathcal{T}_{-}\) both vanish, using the relation \(\rho \p\varphi=-\frac 1{2\pi\beta}\nabla_{\bar{z}} [ ( \partial \varphi)^2]\) we
obtain \begin{align} \la{loop_diff1}
\partial \langle  \varphi \rangle - \partial W   +   \frac{1}{2}(2 - \beta) \partial \log  \left( \sqrt{g} \langle\rho \rangle\right) = \frac{1}{2\pi \beta  \langle \rho \rangle} \nabla_{\bar{z}} \langle ( \partial \varphi)^2 \rangle_{c}.
\end{align}
   Finally, applying $\nabla_{\bar{z}} $ again results in
\begin{align}\la{loop_diff2a}
 \langle \rho\rangle  =   - \frac{1}{4\pi \beta} \Delta_{g} W +  \frac{(2-\beta)}{8\pi \beta} \Delta_{g}\log \Big(\sqrt{g} \langle \rho \rangle\Big) - \frac{1}{2\pi^{2} \beta^{2} } \nabla_{\bar{z}}\left( \langle \rho \rangle^{-1} \nabla_{\bar{z}} \langle ( \partial \varphi)^2 \rangle_{c}\right).
\end{align}
In this form, we can specify the source potential $W$ and develop an asymptotic expansion of the density in magnetic length. Using $W = - K/2l^{2} + (1 - s) \log \sqrt{g}$, we obtain
\begin{align}\la{loop_diff2}
 \langle \rho\rangle  =  \frac {1}{ 2\pi \beta l^{2}} + \frac{ (\beta - 2s)}{8\pi \beta} R +  \frac{ (2-\beta)}{8\pi \beta} \Delta_{g}\log \langle \rho \rangle - \frac{1}{2\pi^{2} \beta^{2} }\nabla_{\bar{z}}\left( \frac{\nabla_{\bar{z}}\langle ( \partial \varphi)^2 \rangle_{c}}{\langle \rho \rangle} \right).
\end{align}
In this  form the Ward identity is the most suitable for the   gradient expansion of the density $\rho$, or equivalently, the expansion in orders of $l^{2}$.   

\subsection{Structure of the Asymptotic Expansion} 
 The Ward identity allows us to compute correlation functions, such as those of the particle density, in the  small $l^{2}$ limit. We can check the identity by integrating both sides of \eq{loop_diff2} to obtain the global relation \eq{shift-spin}. Upon integration of \eq{loop_diff2}, only the first two terms on the right hand side are non-zero. The Gauss-Bonnet theorem
yields \eq{shift-spin} and correctly fixes $N$ to the critical value such that the electronic droplet covers the entire surface. Indeed, such an expansion in terms of curvature invariants is only possible for this critical value of $N$ when the droplet covers the entire surface.

The advantage of  the Ward identity is that it contains terms of different
order in powers of \(N_\phi \) or equivalently in powers of magnetic length.
 Iterating this
equation on a curved surface yields the gradient expansion in   curvature.
The first two terms on the
right hand side of \eq{loop_diff2} capture  the expansion of the density up to order \(l^0\). This is a classical result that misses the effects due to quantum fluctuations, which first appear at order $l^{2}$. To compute it, we must evaluate
the two-point function $\lim_{\xi \to \xi'} \langle (\partial \varphi(\xi)\partial \varphi(\xi')\rangle_{c}=\langle (\partial \varphi)^{2}\rangle_{c}$ at coincident points. Due to the apparent divergence of the correlation function as we merge points,  we must regularize it, which is the subject of the next section.

 The structure of the gradient expansion is similar to  the asymptotic expansion of the Bergman kernel
at coincident
points  which is valid for IQH states where $\beta=1$. The Bergman kernel is\begin{align}
B(\xi,\xi')=\sum_{k=0}^{N_1}\overline{\psi_k(\xi)}\psi_k(\xi'),
\end{align}
where \(\psi_k\) is the one particle wave function \eq{2.7}. At equal points,
the Bergman kernel is the density of the IQH ground state at filling fraction $\nu = 1$, namely \(\langle \rho(\xi)\rangle=B(\xi,\xi)\).

 The expansion of the Bergman kernel, developed in  \cite{zelditch2000szego, lu2000lower,catlin1999bergman, xu2012closed,Klevtsov2013} (an accessible review can be found in \cite{Douglas2010}), shows that at every order in $l$ (or equivalently $N_{\phi} = V/2\pi l^{2} $), the expansion is written in terms of local,  invariant geometrical quantities. The first few terms of the expansion read \begin{align}\la{5.19}
\langle \rho\rangle = &\frac{1}{ 2\pi l^{2}}+ \frac 1{8\pi} R+\frac{ l^2}{3}\frac 1{8\pi}\Delta_{g} R+\frac{l^4}{16\pi}\left(\frac
18\Delta_{g}^{2} R-\frac 5{48}\Delta_{g} (R^2)\right)+\\&\frac{l^{6}}{32\pi}\left(\frac{ 29}{720}\Delta_{g}
(R^3)-\frac{7}{160}\Delta_{g}^2(R^2)+\frac 1{30}\Delta_{g}^{3} R-\frac{11}{120}\Delta_{g}(R\Delta_{g}
R)\right)+\dots\nn
\end{align} 
In the FQH case, the structure of the expansion has the same form, but the coefficients also depend on the characterization of the state - the filling fraction $\nu$, shift $\mathcal S$, and gravitational anomaly $\kappa$. At the present stage the coefficients of  terms
of orders higher than \(l^2\) for the FQH case are not known.

\section{Short Distance Regularization and the Gravitational Anomaly}\la{UV_reg}
\subsection{Short Distance Regularization}
 In this
section we compute the connected function \(\langle(\p\varphi)^2\rangle_c \)   to the leading order,  which allows us to evaluate the Ward identity.  We show that although the function is composed entirely of holomorphic components its expectation value is no longer holomorphic in curved space and for this reason refer to it as the gravitational anomaly.

 We obtain the two-point function from \eq{rhoW} by varying the one-point function of \(\langle \varphi \rangle\) with respect to \(W\). To do this,  the variations must be taken at arbitrary $W$. After the variation,  $W$ is fixed to its value \eq{141} or \eq{2.17}. From Eq. \eq{loop_diff2a}, we get (up to an unknown constant)
\begin{align}\la{6.1}
\langle \varphi \rangle = W + O(l^{2}).
\end{align}
 A variational derivative  yields the connected function
\begin{align}\la{phi_rho_1}
 \langle \varphi(\xi) \rho(\xi') \rangle_{c} =\frac{1}{\sqrt{g(\xi')}} \frac{\delta \langle \varphi (\xi)\rangle}{\delta W(\xi')}     = \frac{1}{\sqrt{g}} \delta^{(2)}(\xi - \xi')-\frac 1V,
\end{align}
or equivalently 
\begin{align}\la{diff_green}
\Delta_{g}\langle \varphi (\xi)\varphi(\xi')\rangle_c =  -4\pi \beta \Big[ \frac{1}{\sqrt g}\delta^{(2)}(\xi-\!\xi')-\frac 1V\Big].
\end{align}
The constant offset \(V^{-1}\) is required by the condition
 \(
\int  \langle \varphi(\xi) \rho(\xi') \rangle_{c} dV'= 0
\),
 which holds identically for a connected correlation function of the density.
%

Proceeding with \eq{diff_green}, we see that the connected correlation function of the field \(\varphi\)
is the Green function  $G$ of the Laplace-Beltrami operator 
\begin{align}\la{phi_green}
\langle \varphi(\xi) \varphi(\xi')\rangle_{c} = 4\pi \beta G(\xi, \xi'), \end{align}
defined as 
\begin{align}\la{diff_green1}
-\Delta_{g}G(\xi,\xi')=  \frac{1}{\sqrt g}\delta^{(2)}(\xi-\!\xi')-\frac 1V.
\end{align}
 However, this and \eq{diff_green} are valid only for  separations
larger than magnetic length.
At short distances corrections are not perturbative in \(1/N_\phi\) and   the leading approximation \eq{phi_green} fails. Indeed, when merging points
the Green function diverges logarithmically. We must therefore regularize it and remove  the singularity  in a manner that retains  general covariance. In the leading order, the only possibility is to subtract off the logarithm of the geodesic distance  $d(\xi, \xi')$ between points (measured in units of magnetic length)
\begin{align}\la{green_reg}
\lim_{\xi \to \xi'} \langle \varphi(\xi) \varphi(\xi')\rangle_{c} = 4\pi \beta G^R(\xi),\end{align}
where 
\begin{align}G^R(\xi)=\lim_{\xi \to \xi'} \left[G(\xi, \xi') + \frac 1{2\pi}\log \frac{d(\xi, \xi')}{l}\right]\la{green_reg2}\end{align}
is the covariant regularized Green function. A more careful treatment would instead use the spatially varying length scale constructed from the particle density $l \to 1/\sqrt{\rho(\xi)}$, which to leading order is just proportional to the magnetic length.

 Expanding the geodesic distance at small separation yields (see e.g. \cite{Ferrari2012a})
\begin{align}
2 \log d(\xi, \xi') & =  \log |z - z'|^{2} +\log\sqrt g+\frac 12(z' - z)\partial \log\sqrt g+ \frac 12(\bar{z}' - \bar{z}) \bar{\partial} \log\sqrt g\\ \nn
&- \frac{1}{12} |z' - z|^{2} |\partial \log\sqrt g |^{2} + \frac{1}{48}\left[(z' - z)\partial \log\sqrt g + (\bar{z}' - \bar{z}) \bar{\partial}\log\sqrt g\right]^{2}\\ \nn
& + \frac{1}{6} \left[(z' - z)^{2} \partial^{2} + 2 |z' - z|^{2} \partial \bar{\partial} + (\bar{z}' - \bar{z})^{2} \bar{\partial}^{2}\right] \log\sqrt{g} \,   + ...\la{geo_1}
\end{align}
The asymptotic behavior of the Green function at coincident points is 
\begin{align}
G(\xi, \xi') \to - \frac 1{2\pi}\log |z - z'|.
\end{align}
The first term from the expansion of the geodesic distance cancels the divergent part of the Green function.  Thus, in the leading order, $\langle \varphi^{2} \rangle_{c}\approx 4\pi\beta G^R $ is finite. A more detailed formula  (see  \eq{G_reg} in \ref{G}) for the regularized Green function implies 
\begin{align}\la{phi_green1}
\langle \varphi^{2}\rangle_{c} \approx 4\pi \beta G^R(\xi)=\beta \log\sqrt {g(\xi)}+\frac{2\pi \beta}{V} K(\xi) - 4\pi \beta A^{(2)}[g, g_0] -\beta,
\end{align} 
where the metric dependent constant $A^{(2)}$ is given by \eq{genfun1}, and $g_{0}$ is the metric of the sphere. 
The regularization procedure is equivalent to the familiar heat-kernel regularization. The value of $G^R$ determines the gravitational anomaly, which in its turn, controls the transport coefficients. The regularized Green function is also a crucial ingredient in the adiabatic phase of motion of quasi-holes  (see \ref{Q1}).  For an infinite plane or a uniform sphere, $G^R$ is a constant and the gravitational anomaly is difficult to see. 
\subsection{ Gravitational Anomaly}
The connected correlated function of $\partial \varphi$ must follow from the short distance regularization \eq{green_reg}
\begin{align}
  \langle (\partial \varphi)^{2}\rangle_{c} =4\pi \beta \lim_{\xi \to \xi'}\partial_{z}\partial_{z'}\left(G (\xi,\xi')+ \frac 1{2\pi} \log \frac{d(\xi,\xi')}{l}\right).
  \end{align}
From the expansion \eq{geo_1}, we find
\begin{align}
\lim_{\xi \to \xi'}\partial_{z} \partial_{z'} (2 \log d(\xi,\xi')) & =  \frac{1}{(z
- z')^{2}} + \frac{1}{6} \left( \partial^{2}\log \sqrt{g}- \frac{1}{2}(\partial \log \sqrt{g})^{2}\right).\la{geo_2}
\end{align}
The leading singularity here removes the divergence of the Green function $\partial_{z} \partial_{z'}G = -(1/4\pi)(z - z')^{-2}$, and we find 
\begin{align}
  &\langle (\partial \varphi)^{2}\rangle_{c}  = \frac \beta 6 \left( \partial^{2}\log \sqrt{g}- \frac{1}{2}(\partial \log \sqrt{g})^{2}\right)
  \la{6.7},
 \end{align}
 is the the Schwarzian derivative. It is  invariant under M{\"o}bius transformations.  Furthermore,
 \begin{align}
&\nabla_{\bar{z}} \langle (\partial \varphi)^{2}\rangle_{c} = - \frac{\beta}{24}   \partial R,\quad \nabla_{\bar{z}} = \frac{1}{\sqrt{g}} \bar{\partial}\la{6.8a}\\
&\nabla_{\bar{z}}^{2}\langle (\partial \varphi)^{2}\rangle_{c}  = - \frac{\beta}{96} \Delta_{g} R ,   \la{6.8}.
\end{align}
The various forms of the connected two-point function provide a  physical basis for understanding its short distance behavior. We find that the correlator \eq{6.7} is proportional to the Schwarzian derivative. Moreover the connected two point function  \eq{6.7} can be recast as $\nabla_{\bar{z}}^{2}\langle (\partial \varphi)^{2}\rangle_{c}$,  a coordinate invariant quantity proportional to the Laplacian of the scalar curvature of the background surface. 

This analysis tells us that, to leading order in $l$, the field $\varphi$ can be viewed as a Gaussian free field. As such, a part of the holomorphic component of the stress-energy tensor $ -(\partial \varphi)^{2}$ is not  actually holomorphic
unless the space is flat. This observation, captured by formulas \eq{6.7}-\eq{6.8}, is often referred as the gravitational anomaly or trace anomaly (see e.g., \cite{Polyakov1987}). For a 2D Coulomb plasma, a model of statistical mechanics which is equivalent to the Laughlin states,  the effects of short distance regularization were first outlined in the important early papers Ref. \cite{Jancovici, Jancovici1}


We emphasize that all these formulas hold only in the leading approximation
in the gradient expansion. In particular, \eq{6.7} acquires corrections proportional to the curvature. These corrections, although interesting, occur in higher orders of $l^{2}$ than we consider in this paper. We will proceed to explore some of the consequences of \eq{phi_green}, \eq{phi_green1}, and \eq{6.7}. In \ref{shortd}, we give an alternative derivation of the short distance regularization of \eq{green_reg}.

\section{ Gradient Expansion}\la{Sec7}
Our formalism allows us to compute the gradient expansion of the density beyond the Eq.\eq{1}. We use the covariant regularization of the correlation function \eq{6.7} to convert the Ward identity  (\ref{5.8},\ref{5.81}) into a closed equation written in terms of one-point functions. 

Using \eq{6.7}
we write the Ward identity \eq{5.8}  in the leading \(O(1/N_\phi)\) order  {  
 \begin{align}\la{7.1}
-\langle\mathcal{T}_-\rangle&=\int \frac{  \partial W}{z - \xi}  \langle\rho\rangle
  dV  +\frac{1}{2\beta}(\partial \langle\varphi   \rangle)^2+ \frac{2 - \beta}{2\beta}
  \partial^2\langle\varphi\rangle+\\& + \frac{1}{12} \left[\partial^{2} \log \sqrt{g}
- \frac{1}{2}\left( \partial \log \sqrt{g}
\right)^{2}
\right]=0.\nn
\end{align}
}
For solution of this equation, the Ward identity in the form \eq{loop_diff2}
is more convenient. To the leading order, and by virtue of \eq{6.8a}, we  write
\begin{align}
\nabla_{\bar{z}}\left( \frac{1}{\langle \rho \rangle}\nabla_{\bar{z}} \langle (\partial \varphi)^{2}\rangle_{c} \right) =  - \frac{\pi \beta^{2} }{48 } l^{2}\Delta_{g}R + O(l^{4})
\end{align}
and substitute it into  \eq{loop_diff2} \begin{align}\la{7.2}
 \langle \rho\rangle  =    \frac{1}{2\pi \beta l^{2}} + \frac{ (\beta-2s)}{8\pi \beta } R +  \frac{(2-\beta)}{8\pi \beta} \Delta_{g}\log \langle \rho \rangle +  \frac{ 1}{96\pi } (l^{2}\Delta_{g})R + O(l^{4}).
\end{align} 
Iterating this expression and matching terms of equal order yields the formula \eq{rho}. For general spin $s$
the formula reads\begin{align}\la{density_spin}
\langle \rho\rangle  &= \frac{1}{2\pi \beta l^{2}}+ \frac{ \beta-2s}{8\pi \beta }  R + \frac{b}{8\pi} ( l^{2} \Delta_{g} )R,  \qquad \, b =\frac{1}{12} + \frac{2 - \beta}{4\beta} (\beta-2s).
\end{align}
Computing next terms of the gradient expansion requires further details
in the short distance regularization. 
\section{Transport Coefficients for the Laughlin State}

Now we can compute the gradient expansion of  transport coefficients listed in the introduction (again we measure momentum $k$ in units of $l^{-1}$).

 Eq. \eq{density_spin}  determines the response to curvature $\eta(k)$, 
\begin{align}\la{eta_full}
\eta(k) = \frac{\beta - 2 s}{4} - \frac{ b}{4 \nu } k^{2} + O(k^{4}),
\end{align}
and the relation \eq{eta_sk} determines the static structure factor up to order $k^{6}$ \eq{1.8}. In equivalent form it is
\begin{align}\la{static_structure_factor}
S(k) = \frac{1}{2} k^{2} + \frac{\beta - 2}{8} k^{4} + \frac{(3 \beta - 4)(\beta - 3)}{96} k^{6} + O(k^{8}).
\end{align}
 As expected it is independent of spin $s$. We notice  that our calculation of the $k^{6}$ coefficient agrees with a computation of the same coefficient but by a completely different approach in \cite{Kalinay2000}. In this reference, the authors developed a diagrammatic Mayer expansion for the log-plasma in a neutralized
background. At the moment, the exact determination of the static structure factor  beyond $k^{6}$ is known only for the
integer case, for which $S(k) = 1 - \exp(- k^{2}/2)$. This determines the coefficients of the terms in  \eq{5.19} linear in $R$ through the relation \eq{eta_sk}.

Having the structure factor we obtain the momentum dependence of the Hall
conductance by virtue of \eq{3.15}.  It also establishes a relation between the momentum dependence of the Hall conductance and the response to the curvature and the anomalous
 viscosity  $\eta^{(A)} $(see \ref{B1}) (cf., \cite{Hoyos2012}).
 
 We emphasize that all  corrections but
the gravitational anomaly  to the Hall conductance vanish for the bosonic Laughlin state. The Hall conductance misses the \(O(k^2)\) correction \begin{align}\la{7.6}
\sigma_H(k) = \sigma_H(0) \left ( 1 - \frac{ k^{4} }{24}
 + O(k^{6})\right),\quad \nu=1/2.
\end{align}
The Hall current decreases with a long-wave modulation. 

On the other hand, the Laughlin 1/3 states misses the \(O(k^4)\) correction.
In this case the gravitational and mixed anomalies accidentally cancel each
other \begin{align}\la{7.61}
\sigma_H(k) = \sigma_H(0) \left ( 1 + \frac{k^{2} }{4}
 +O(k^{6})\right),\quad \nu=1/3.
\end{align}  
In this case long-wave modulations enhance  the Hall current.

\section{Generating Functional}\la{Sec9}
\subsection{The Structure of the Large $N_\phi$ Expansion and Geometric Functionals}\la{9.2}
Once we know the \(l^{2}\) expansion of the density \eq{density_spin}, the expansion of the  generating functional  is obtained by integrating \eq{3.7} \begin{align} \la{9.1}
 - \frac 12 l^2\Delta_g\frac {\delta \log{\mathcal Z}[g]} {\delta\sqrt g}&=\Big ( 1 -\frac{ ( 1 - s)}{2} {l^2} \Delta_g\Big)\<\rho\>.
\end{align} 
This expansion produces covariant  functionals which depend only on the geometry of the manifold and also a sequence of coefficients which depends only on characteristics of a  QH state, such as the filling fraction, shift, and gravitational anomaly. These functionals  recently came to a prominence in K\"ahler geometry.

 Integrating \eq{9.1} produces an asymptotic expansion of the generating functional in the small dimensionless parameter $1/N_\phi=2\pi l^{2}/V$. Using \eq{density_spin}, we find the first three terms of the expansion for Laughlin states 
\begin{align} \nn
\log\frac {{\mathcal Z}[g]}{{\mathcal Z}[g_0]} 
=-\frac{2\pi}{ \beta} N_{\phi}^{2}A^{(2)}[g,g_0]+\frac{ (\beta-2s)}{ 4\beta } N_{\phi}A^{(1)}[g,g_0] -\frac{1}{96\pi}& \left(1- 6\frac{ (\beta - 2s)^{2}}{2 \beta}\right) A^{(0)}[g,g_0]+
\\& + \sum_{k\ge  1} p_{-k}N_{\phi}^{-k} A^{(-k)}. 
\la{logZ}
\end{align}
Here, $A^{(k)}[g,g_0]$ are functionals of equal area metrics $g$ and $g_{0}$ in the same conformal class. The leading three orders are given by
\begin{align}
&A^{(2)}[g,g_{0}]   = \frac{1}{4 V^{2}} \int K dV - K_{0} dV_{0},\la{genfun1}\\
& A^{(1)}[g,g_{0}] = \frac{2}{V}\int \log \sqrt{g} dV - \log\sqrt{g_{0}} dV_{0},\la{genfun2}\\
&A^{(0)}[g,g_{0}]   = \int R \log \sqrt{g} dV - R_{0} \log \sqrt{g_{0}} dV_{0},\la{genfun3}
\end{align}
where $dV_{0} = \sqrt{g_0} d^{2} \xi$. This form, although transparent,  does not have a rigorous meaning since individual terms in this formula are not defined on the entire surface, whereas the difference which appears in the integrand is globally defined. To emphasize this point it is customary to express the functionals only in terms of globally defined functions $2u=K-K_0$ and $2\sigma=\log(\sqrt g/\sqrt{g_0})$ (\ref{gs},\,\ref{ks}).  They can be found in \cite{Ferrari2012a}. For reference, we present them in \ref{largeN}.

 The geometric functionals are invariant characterizations of the manifold. They do not depend on the choice of coordinates. This fact is not obvious from the explicit form (\ref{genfun1}-\ref{genfun3}). The best way to see it is to come back to their origin. In complex geometry, these geometric functionals appear as a result of the transformation properties of the Green function for the Laplace-Beltrami operator. See for example \cite{Ferrari2012a}.  For reference, we list these transformation formulas in \ref{G}.

 The functional \({A}^{(2)}\) is known in the literature on K\"ahler geometry as the Aubin-Yau functional. The shifted functional \( A_{M}^{(1)}={A}^{(1)} + 4\pi \chi {A}^{(2)}\) is known as the Mabuchi or K-energy functional (see e.g. \cite{Ferrari2012a} and references therein). 

The functional  \({A}^{(0)}\) is Polyakov's Liouville action \cite{Polyakov1987}.  It is a logarithm of the spectral determinant of the Laplace-Beltrami operator. It represents the effect of the gravitational anomaly.  Normalizing with respect to a reference metric it reads
 \begin{align}\la{P}
-\frac{1}{2}\log\frac{{\rm Det}\left(-\Delta_g\right)}{{\rm Det}\left(-\Delta_{g_0}\right)}=\frac{1}{96\pi} A^{(0)}[g,g_0].
\end{align} 

\subsubsection*{Cocycle Condition and Higher Order Functionals}
The leading order expansion of the generating functional is distinguished from higher orders by the cocycle condition. For a triple of metrics $g$, $g_{0}$, and $g_{1}$ in the same conformal class, the cocycle condition is defined by the relation
\begin{align}\la{cocycle}
A^{(k)}[g, g_{0}] = A^{(k)}[g, g_{1}]  + A^{(k)}[g_{1}, g_{0}].
\end{align}
A functional is called an {\it exact one-cocycle} if it can be written as $A^{(k)}[g, g_{0}] = A^{(k)}[g] - A^{(k)}[g_{0}]$, with $A^{(k)}[g]$ a local integral of the curvature $R$ and its covariant derivatives. Exact one-cocycles trivially satisfy the cocycle condition. 

The LHS of \eq{logZ} clearly satisfies \eq{cocycle}, and since the functionals occur at different orders in $N_{\phi}$, they must satisfy \eq{cocycle} individually. While the geometric functionals written in the form \eq{genfun1}-\eq{genfun3} do also appear to trivially satisfy \eq{cocycle}, they are not exact one-cocycles since each term taken  separately  is not well defined.  Nevertheless, they do satisfy the co-cycle property. These functionals are distinct from those which appear at higher order $k \ge 1$: they are not exact one-cocycles, and for that reason  cannot be expressed locally through the scalar curvature $R$ and its covariant derivatives. The terms which appear at higher orders in $1/N_{\phi}$ are exact one-cocycles, and can be expressed as the difference of local functionals of the curvature for each metric.  These cocycle properties of the expansion were first conjectured for the integer $\beta =1$ case in \cite{Klevtsov2013}, and later shown to hold for all $\beta$ in a \cite{KlevtsovGFF}. For example, 
\begin{align}\la{A-1}
A^{(-1)}[g] = V \int R^{2} dV.
\end{align}
 At order $1/N_{\phi}$, this is the only local functional of the curvature available. Higher order corrections have a more complicated structure, but they are also local. The functionals for the integer case $\beta = 1$ have been computed through the asymptotes of the Bergman kernel \cite{Klevtsov2013}. The next order which appears for $\beta = 1$ is

\begin{align}\la{A-2}
A^{(-2)}[g] =V^{2}\int (29 R^{3} - 66 R \Delta_{g} R ) dV,\quad (\beta = 1)
\end{align}
We give more details and summarize some known results at $\beta = 1$ in \ref{largeN}.


\subsection{Generating Functional for Density Correlation Functions}
It is often more useful to consider variations of the general potential  \(W=-K/2l^{2}+ (1-s)\log\sqrt g\) than variations over the metric.  At $s = 1$, the source potential $W
= - K/2l^{2}$ is directly proportional to the K\"ahler potential.
This helps to  write the density  as a functional of $W$ by setting $s=1$ in formulas \eq{density_spin} and replace $R$  in \eq{density_spin}   by $,\, 2(l^2\p \bar{\p}W)^{-1}\p\bar\p \log(-\p\bar\p W)$, 
\begin{align}\la{8.2}
{  \sqrt g}\langle \rho \rangle\! =&-\frac{\p\bar\p
W}{\pi \beta } + \frac{(2-\beta )}{2\pi
\beta} \p\bar\p \log \left( -\p\bar\p
W \right) \\& +\frac{1}{ 12\pi}\left(1
\!-\!
 6\frac{(2-\beta)^{2}}{2\beta}\right) \p\bar\p\left[
 \frac{1}{ \p\bar\p
W} \p\bar\p \log (-\p\bar\p
W )\right].\nn
\end{align}
 Similarly the \(N_{\phi}\) expansion of the generating functional reads as an expansion in gradients of the function $W$. It can be obtained  in a similar manner by setting $s=1$ in  \eq{logZ} and replacing $K$ and
 $\sqrt g$  in  (\ref{genfun1}-\ref{genfun3}) by $-2l^2W$ and $-2l^2\p\bar\p W$, respectively.  As before, the expansion can be written as $\log \mathcal{Z}[W]/\mathcal{Z}[W_{0}]$, where ( up to a metric independent terms) 
\begin{align}
\log \mathcal{Z}[W]  &=  \frac{1}{2\pi \beta} \int\bar\p W\p W  d^{2} \xi + \frac{2 - \beta}{2\pi \beta} \int (\p\bar\p W) \log(- \p\bar\p W) \, d^{2} \xi\la{8.3}\\
&   + \frac{1}{24\pi} \left( 1 - 6\frac{ (2-\beta )^{2}}{2\beta}\right) \int \log (-\p\bar\p W) \p\bar \p \log (-\p\bar\p W) d^{2} \xi.\nn
\end{align}
%
%
%
We must also chose the reference function $W_{0}$ appropriately in order for the generating functional $\log \mathcal{Z}[W]/\mathcal{Z}[W_{0}]$ to be well defined.

 This formula generates
multiple correlation functions of the density and also describes the thermodynamics
 of the 2D Dyson gas. It has been previously
obtained in \cite{Zabrodin2006}. The difference here is that we do not consider droplets with a boundary, and consequently (\ref{8.3}) should be compared with the bulk terms obtained in \cite{Zabrodin2006}.

\subsection{Generating functional in an inhomogeneous magnetic  field} The structure of the formula \eq{8.3} is further clarified if we  assume that
magnetic field is not homogeneous. To obtain it we  express $\p W$ in terms of the vector
potential  $\sqrt g B=2\ii(\bar\p A-\p\bar A)$ and the spin connection $\frac
12\sqrt g R=2\ii(\bar\p \omega-\p\bar \omega)$. In the transversal gauge $\bar\p A+\p\bar A=\bar{\partial} \omega+\p\bar\omega=0$ and in units $e = \hbar = 1$, 
  \begin{align}\p W=-2\ii \left[ A+(1-s)\omega\right], \quad \p\bar\p W=  -\frac 12\sqrt g\mathcal B,\la{131}\end{align}
where we denoted  
 \begin{align}\mathcal{B}=B+\frac 12(1-s) R.\la{132}\end{align} 
Then \eq{8.3} reads 
\begin{align}
\nn\log \mathcal{Z} =& \frac{2}{\pi} \int \left[ \frac{1}{\beta} \left(A   +
 \frac{1}{2} (\beta - 2s)  \omega\right)\left( \bar{A}  + \frac{1}{2}(\beta - 2s) \bar \omega \right)
- \frac{1}{12} \omega \bar{\omega}
\right]d^2\xi\\
&-\frac 1{16\pi\beta}\int\left[\left(\frac \beta 3-(\beta-2)^2\right)\left( R-\frac 12\Delta_g\log
 \mathcal{B}\right)-4(\beta-2)\mathcal{ B}\right]\mathcal{\log B\,} dV.
\la{8.31}
\end{align}

The first line of this formula 
 represents the  anomalies: the  gauge anomaly ($A\bar A$ term),  the mixed anomaly ($A\bar \omega+\bar A\omega$ term) and the gravitational anomaly ($\omega\bar\omega$ term).
The second line is local. It is expressed  in terms of magnetic field and the curvature.

Among the applications of the formula \eq{8.2}  is the thermodynamics of the 2D Coulomb plasma. We discuss it in \ref{HH}.  Another application is the effect of a inhomogeneous magnetic field.

\section{Transport Coefficients in an Inhomogeneous Magnetic Field and a Curved Space}\label{E}
The  formula \eq{8.2} gives the density in a non-uniform magnetic field. Using (\ref{131},\ref{132}) we obtain   
 \begin{align}\nn
\langle \rho \rangle =  \frac{ \nu B }{2\pi}+  \frac{\beta - 2s}{8\pi \beta} R + \frac{2 - \beta}{8\pi \beta}
\Delta_g \log \mathcal{B} + \frac{1}{8\pi} \left( \frac{1}{12} -  \frac{(2
- \beta)^{2}}{4\beta} \right) \Delta_g \left( \frac{1}{\mathcal{B}} (R-\Delta_g \log \mathcal{B}) \right)
+ \dots,
\end{align}
where $\mathcal{B}$ is given by  \eq{132}. The gradient expansion of the Hall conductance \eq{7.6} follows from this
formula by virtue of \eq{response_B}.  
The second term in this expansion (in the flat case at $R=0$) was previously obtained in \cite{SonNC, Wiegmann2013}.  As before, for the bosonic Laughlin state $\nu=1/2$ (or more generally $\mathcal{S}=2$) this term vanishes. The gradient corrections are then solely due to the gravitational anomaly. For small deviations from a uniform magnetic field, and the flat space the density is 
 \begin{align}\nn \text{Bosonic Laughlin state}:\qquad 
\langle \rho \rangle =  \frac{{\nu B}}{2\pi}-\frac{1}{12}\frac{ \Delta B}{8\pi B }   +\dots
\end{align}

In a uniform magnetic and electric fields, this equation  and the St\v reda formula \eq{0} give the correction to the quantized  value of the Hall conductance by gradients of curvature   \be\nn\sigma_{H} =\frac{e^2}h\left( \nu- \frac{b l^{4}}{4} \Delta_{g} R\right), \ee
where $b$ is given by \eq{density_spin}. Curiously, the correction vanishes  at $\nu=1/3$ and $s=1$.

\section{Other FQHE States}\la{Sec10} We were able to compute the gradient expansion
of various objects for the Laughlin series of FQH states. Less is known about the similar gradient expansion for other FQH states and iteration of the Ward identity.  Nevertheless, there is a  conjecture describing how the formulas should
generalize to the states associated with the affine \(\widehat{sl}_k(2)\)
Kac-Moody algebra. These are Pfaffian and parafermionic states \cite{Moore1991,Read99} which, like the Laughlin state, consist of one sort of particle and are therefore characterized by one density
function.   

The conjecture follows from the sum rules \eq{3.9} and the arguments in  \cite{Abanov2013} and \cite{GromovAbanov}. It assumes
 that the form  of the Ward identity \eq{7.1} remains the same except the
coefficients in front of the last two terms on the RHS   {  
\begin{align}\la{9.20}
-\langle{\mathcal{T}}_-\rangle&=\int \frac{  \partial W}{z - \xi}  \langle\rho\rangle
  dV  +\frac{1}{2}\nu\left[(\partial \langle\varphi   \rangle)^2+ (2 - \mathcal{S})
  \partial^2\langle\varphi\rangle\right]+\\& - \frac{\kappa}{12} \left[\partial^{2} \log
\sqrt{g}- \frac{1}{2}\left( \partial \log \sqrt{g}\right)^{2}
\right]=0.\nn
\end{align}
}
The parameters \(\nu \) and \(\mathcal{S}\) are the filling fraction and shift, respectively, defined by \eq{2.23}, and {  \( \kappa \)} is a new parameter. For the Laughlin state {  \(\kappa=-1\)} and \(\mathcal{S}=\nu^{-1}\).  For the $\mathbb{Z}_{k}$ parafermion state, discussed in \ref{A}, the shift $\mathcal{S} = M + 2$, and the filling fraction $ \nu = k / (M k + 2)$, for integer $M$.

 The coefficients in front of \((\partial \langle\varphi   \rangle)^2\) and \(
  \partial^2\langle\varphi\rangle\) are consistent with the  sum rule \eq{3.9}
  and the global relation \eq{shift-spin}. We  see this by computing the gradient expansion of the density starting
from \eq{9.20}. Assuming \eq{9.20} and repeating the calculations  of the Sec. \eq{Sec7} we obtain  {  
\begin{align}\la{p}
&\langle \rho\rangle = \frac{\nu }{2\pi l^{2}} +  \frac{ \nu (\mathcal{S}-2s) }{8\pi} R
+ \frac{b}{8\pi}  l^{2}\Delta_{g} R + O(l^{4}), \\ &b = -\frac{\kappa}{12} +  \frac \nu 4 ( \mathcal{S}-2s)(2 - \mathcal{S}),
\end{align}
} 
where the first two terms are  the same as in \eq{shift-spin} for all FQHE states. The generating functional \eq{gen_fun} for this state is presented in \ref{largeN}.  Here we present only the anomalous part of the functional (the generalization of the  first line of \eq{8.31}) 
\begin{align}
\log \mathcal{Z} =& \frac{2}{\pi} \int \left[ \nu \left(A  +
 \frac{1}{2} (\mathcal{S} - 2s) \omega \right)\left( \bar{A}  + \frac{1}{2}(\mathcal{S} - 2s) \bar\omega\right)
+ \frac{\kappa}{12} \omega \bar{\omega}
\right]d^2\xi
\la{8.311}
\end{align}

 The value of {  $\kappa $}  is conjectured  to be {  {\it minus}} the central charge of the  \(\widehat{sl}_k(2)\)
Kac-Moody algebra {  \(\kappa =-\frac{3k}{k+2}\)}. In particular, the Pfaffian state corresponds to {  $k=2,\, \kappa =-3/2$.}
 Although it seems plausible, we emphasize that to the best of the authors' knowledge, this conjecture has not been obtained by calculations similar to those presented here.    

 The formulas \eq{eta_sk} and \eq{3.15} which connect the density response
 to curvature,  the Hall conductance, and
the static structure factor  hold for all FQHE states in the lowest Landau level. For example, the  static structure factor 
reads  
 \begin{align}
  S (k)    =\frac{k^{2}}{2}  + (\mathcal{S}- 2)\frac {k^4}8 +\left[3 (2 - \mathcal{S})^{2} +\kappa \nu^{-1}\right] \frac{k^6}{96} + O(k^{8}).
  \end{align}
The appearance of the gravitational anomaly in the static structure factor means that it is in principle accessible to numerical investigation. For example, for the bosonic  Pfaffian state at $\nu = 1$, $\mathcal{S} = 2$, {   $\kappa =-3/2$}, as well as for the bosonic Laughlin case $\nu = 1/2$, $\mathcal{S} = 2$, {   $\kappa =-1$} the leading $k^4$ term in the long-wave expansion of $S(k)$ vanishes. The sub-leading non-vanishing term is controlled solely by  the gravitational anomaly { 
\begin{align}
 S(k) = \frac{k^{2}}{2} +\kappa \frac{k^{6}}{96 \nu} + ... \left\{
  \begin{array}{lr}
\kappa  = -3/2,\, \nu = 1 \quad\text{Bosonic Pfaffian state}, \\
\kappa = -1, \, \nu = 1/2\qquad \text{Bosonic Laughlin state}. 
  \end{array}
\right.
\end{align}}
A computation of the sixth moment sum rule for this state would then determine {  $\kappa$.}

 \section{FQHE as the Gaussian Free Field}\la{4}

 In this section we formulate the generating functional  for the Laughlin states as a field theory of a  Gaussian Free Field.  The formulation involves the measure of integration over density. This approach   is justified by the results from the Ward Identity. A similar field theoretic formulation for the flat case with a boundary is found in \cite{Zabrodin2006}.
  
  The advantage of the formulation in terms of a field theory is that it provides a physically motivated short-cut to the results for the generating functional. It also offers a practical approach to extend the results to more general QH states. In this section we limit ourselves to the Laughlin states.
  \bigskip
  
 We attempt to represent observables in QH states as a path integral over the field $\varphi$  defined by \eq{4.4}, or equivalently, over  collective coordinates \eq{a}
 \begin{align}\la{13.1}D\varphi=\prod_{k > 0} da_{-k} d\bar a_{-k}. 
 \end{align} 

 In particular we wish to write the generating functional as a path integral over some field. In this section,  we show that
   \begin{align}\la{13.10}
 \mathcal{Z}[g]={\rm Det}\,(-\Delta_g)\int e^{-\Gamma[\varphi]}D\varphi.
 \end{align}
 We will show that the action  $\Gamma [\varphi, g]$ consists of  distinct parts: the wave function, the entropy $S[\rho]=-\int \rho\log(\rho\sqrt g)\, dV$, and the metric 
 \begin{align}\la{13.2}
 e^{-\Gamma[\varphi]}=|\Psi|^2 e^{S} \prod_{i} \sqrt{g(\xi_{i})}  \end{align}
The two factors ${\rm Det}\,(-\Delta_g)$ and  $e^S$  come from the Jacobian corresponding to the change of variables from $\xi_1,\dots,\xi_N$ to $D\varphi$ in the limit of large $N$,
  \begin{align} 
   \prod_i d^2\xi_i \sim {\rm Det}\,(-\Delta_g)e^{S} D\varphi.\la{13.3}
  \end{align}
\subsection{Stress Tensor,    Effective Action, Energy and Entropy}
We  show that the holomorphic component of the stress energy is related to  the ``action" in \eq{13.2} by the general conservation
law  valid for all QH states 
{   \begin{align}\la{c}
-\bar\p\mathcal{T}_{-} + \pi \sqrt{g} \rho \partial_z\left(\frac{1}{\sqrt{g}}\frac{\delta \Gamma}{\delta\rho}\right)=0.
\end{align} 
}
The generality of this law suggests  a practical means for extending our results to other QH states. To prove \eq{c}, we apply the  anti-holomorphic derivative
to the projected stress-tensor \eq{set_projection}, use
 the \(\p\)-bar
formula $\bar{\partial}( \frac{1}{z} )= \pi \delta^{(2)}(\xi)\) and pass
from the particle density operator to its expectation value  via the identity \(\sum_i
 \delta^{(2)}(\xi-\xi_i)\partial_{z_i}=\rho(\xi)\sqrt{g}\partial_z \, \frac{1}{\sqrt{g}}\frac{\delta
}{\delta\rho(\xi)} \). Then, we find 
{   \begin{align}
-\bar\p\mathcal{T}_-=&|\Psi|^{-2}\sum_i\pi\nabla_{z_i}  \delta^{(2)}(\xi-\xi_i) |\Psi|^{2}\nn \\
=&\pi\sum_i\left[\delta^{(2)}(\xi - \xi_{i}) \partial_{i} \log \sqrt{g(\xi_{i})}-\p_z\delta^{(2)}(\xi-\xi_i)+
\delta^{(2)}(\xi-\xi_i)\p_{z_i}\log|\Psi|^2\right]\nn\\
=&\pi\sqrt{g}\rho\partial_{z} \frac{1}{\sqrt{g}}\frac{\delta }{\delta\rho}\log|\Psi|^2-\pi \partial_{z} (\sqrt{g} \rho) +\pi \rho \partial \sqrt{g}\nn\\
 =&\pi \sqrt{g}\rho \p_z \left( \frac{1}{\sqrt{g}} \frac{\delta
}{\delta\rho}\log|\Psi|^2-\log \rho\right)= - \pi \sqrt{g} \rho \partial_z\left(\frac{1}{\sqrt{g}}\frac{\delta \Gamma}{\delta\rho}\right),\la{13.6}
\end{align} 
}
where $\Gamma$ is given by \eq{13.2}.
   
For the Laughlin state,  $\mathcal{T}_-$ is known explicitly  \eq{loopint}, and we can therefore determine the action  {  
\begin{align}\la{4.31}
-\Gamma[\varphi]=\beta \int \rho(\xi') \log|\xi - \xi'| \rho(\xi')\,dV dV'+\left(\frac\beta 2-1\right)\int\rho\log (\rho\sqrt g) dV +\int  W\rho\, dV.
\end{align} 
}
Furthermore, we obtain the   field theoretical form of the wave function  via \eq{13.2}.

The action \eq{4.31} has the following interpretation. The probability density of the Laughlin states 
 can be seen as  the Boltzmann weight of the  2D one-component Coulomb plasma (or a 2D Dyson gas) in a neutralizing background  $W$, where the filling fraction \(\nu=\beta^{-1}\) plays the role of the temperature
  \begin{align}\la{E2}
dP=\mathcal{Z}^{-1}\ e^{-\beta E}\prod_i d^2\xi_i,\quad \beta E=-\beta\sum_{i\neq j}\log|z_i-z_j|- \sum_i
W(\xi_i)
\end{align} 
 In the leading orders of the large \(N\) approximation, one may replace the double  sums  by integrals, such that {   \(
-\beta \sum_{i\neq j}\log|z_i-z_j|\to  -\beta \int \rho(\xi)\log|{\xi-\xi'}|\rho(\xi')dVdV' 
\). }
 The double sum in \eq{E2} excludes the term with \(i=j
\) while the integral  does not. To account for this, we must
subtract the term \(\int(\log \ell)\rho dV\) where \(\ell\) is a short distance cutoff reflecting the typical spacing of the particles. It only depends on the metric and particle density. The natural choice is therefore  \(\ell \sim (\rho\sqrt
g)^{-1/2}\).
The energy reads {   \begin{align}
\beta E[\rho]=- \beta \int \int \rho(\xi) \log |\xi - \xi'| \rho(\xi') dV' dV-\frac\beta 2\int\rho\log (\rho\sqrt g) dV-\int W\rho dV.
\end{align}
}
The action is obtained by taking into account Boltzmann entropy. The entropy occurs  when passing from the integration
over coordinates of the individual particles to integration over macroscopic
density:  \(\prod_i d^{2}\xi_i\sim e^{S}D\rho \), where the entropy  is
\(S=-\int \rho\log(\rho\sqrt g)\, dV\).
As a result we obtain \(dP\propto e^{-\Gamma [\rho]}D\rho \), where the effective
``action" is given by \eq{4.31} We observe that the short distance effect and the entropy add together. 
\subsection{Gaussian Free Field,   Background Charge and Measure}
\subsubsection*{Background charge}
So far the action  \eq{4.31} is expressed in terms of the field $\varphi$ (through $\rho$) which is not globally defined on the Riemann sphere.  It is convenient  to express the action in terms of  a  field  defined globally.   To this end we introduce a field $\tilde\varphi$, which is the  solution of the Poisson equation  
\begin{align}\la{13.11}
- \Delta_{g} \tilde{\varphi} = 4\pi \beta \left( \rho - \bar \rho[g]\right).
\end{align}
Here, the density is a source  neutralized by the  background charge 
 \begin{align}\la{13.111}
\bar \rho[g] = 
\frac{1}{ 2\pi\beta l^2} + \frac{ (\beta - 2s)}{8\pi \beta} R,
\end{align}
which saturates  the local version of global relation \eq{shift-spin}. Explicitly 
\begin{align}
\tilde\varphi &= \varphi  - \bar \varphi[g] , \qquad \bar\varphi [g] = -\frac{1}{2l^2} K  + \frac{ (\beta - 2s)}{2} \log \sqrt{g},
\end{align}

In terms of this field, the action \eq{4.31} reads
\begin{align}\la{effective_action}
\Gamma[\varphi]
=&- \frac{1}{8\pi \beta} \int \tilde{\varphi} \Delta_{g} \tilde{\varphi} dV  - \left( \frac{\beta}{2} - 1\right) \int \rho \log (\rho) dV + \Gamma_{c}[g], \end{align}
where the functional 
\begin{align}\la{background_action}
\Gamma_{c}[g]=-\frac{1}{2} \int \bar\varphi [g]\bar\rho [g]dV,
\end{align}

  is the contribution of the background charge to the action.  
 From  general arguments we expect that  such functionals must be expressed in terms of geometric functionals defined in Sec.\ref{9.2}. Indeed, straightforward calculations show
 \begin{align} \la{3.17}
-\Gamma_{c}[g]&+\Gamma_{c}[g_0] =\\& -\frac{2\pi}{ \beta} N_{\phi}^{2} A^{(2)}[g,g_{0}] + \frac{ (\beta - 2s)}{4\beta} N_{\phi} A^{(1)}[g,g_{0}] + \frac{ (\beta - 2s)^{2}}{32\pi  \beta} A^{(0)}[g,g_{0}].\nn
\end{align}
This differs from \eq{gen_fun} only at order $O(1)$. In the next section, we show that the $O(1)$ correction to \eq{3.17} comes from the gravitational anomaly. 

\bigskip

\subsubsection*{Measure}
The
last (and a  subtle)   step is to determine the measure \(D\rho\). 
The relation \eq{13.11}  connects the measure of integrations over density and the field $\tilde\varphi$.  Naturally, the  measure of $\varphi $ and $\tilde\varphi $ are the same.
The Jacobian of this transformation is 
the spectral determinant of the Laplace-Beltrami operator 
 \begin{align}\la{jac}D\rho\sim {\rm Det}( - \Delta_{g})D\varphi={\rm Det}(-\Delta_{g})\prod_{k> 0}da_{-k} d\bar a_{-k}.\end{align} 
 This yields the formula \eq{13.3}. 

 Thus the probability distribution is  \begin{align}
dP=\mathcal{Z}^{-1}[g]\, {\rm Det}( - \Delta_{g})e^{-\Gamma[\varphi]}D\varphi,
\end{align}   
where the generating functional is \begin{align}\la{4.101}
\mathcal{Z}[g]={\rm Det}( - \Delta_{g})\int  e^{-\Gamma[\varphi]}D\varphi.
\end{align}
\bigskip
The integral  \eq{4.101} differs from generating functional of a Gaussian Free Field with the background charge (the second term in the action \eq{effective_action}) by  the \(\int\rho\log \rho\, dV\) term, which only vanishes for the Bosonic Laughlin state \(\beta=2\).  This  term violates the classical conformal invariance otherwise possessed by the Gaussian Free Field.

  \subsection{ Anomalies}
If the gradients of  curvature (measured in units of magnetic length)  are  small,
the integral   \eq{4.101} admits a gradient expansion or, equivalently, an expansion in
\(1/N_\phi\).
To obtain it, we identify the field  $\varphi_c$ which minimize  the action $\Gamma[\varphi_c]$, and then
compute the fluctuation around the extremum. The Gaussian approximation suffices.
It captures all necessary physics we would like to address in this paper. 

The action  has terms of the order \(O(N_\phi^2) \), at 
most. The fluctuations are of the order \(O(1)\).
Therefore we must compute the
classical action (i.e., the value of the action at the saddle point) up to  order  \(O(1)\).  At \(O(N_\phi)\), the  saddle point produces the ``mixed anomaly", while the fluctuations give the gravitational or trace anomaly.

\subsubsection*{Fluctuations: Gravitational Anomaly}
The fluctuations  are given by the determinant
of the Laplace-Beltrami operator. Thus, the generating functional is 
\begin{align}\la{4.10}
\log\mathcal{ Z}[g]=-\Gamma[\varphi_c] +  \frac  12\log{\rm Det}\left(-\Delta_g\right).
\end{align}
Note the Jacobian appearing in \eq{jac} is responsible for flipping the sign in front of the spectral determinant. The value of the spectral determinant is Polyakov's Liouville action \cite{Polyakov1987}. It encodes the  gravitational anomaly and is related to the functional $A^{(0)}$ through Eq.\eq{P}.

\subsubsection*{Contribution of the Background Charge}
The field that minimizes  the action satisfies the Euler-Lagrange equation  
\begin{align}
- \frac{1}{4\pi \beta} \Delta_{g} \tilde{\varphi}_{c} -\frac{1}{4\pi\beta} \left(1-\frac{\beta}{2} \right)\Delta_{g} \log \rho_{c} = 0,
\end{align}
which can be written as a Liouville-like equation with a background 
\begin{align}\la{4.16}
 \rho_c=\bar \rho[g]+\frac{1}{4\pi\beta}\left(1-\frac\beta
2\right)\Delta_g\log\rho_c.
\end{align} 
The terms in this equation are of different orders in $l^{2}$. It must be solved
iteratively up to the order of \(O(l^{2})\), or equivalently to order $1/N_{\phi}$. In the leading approximation we have
 \begin{align}
&\rho_c=\bar\rho[g]+\frac{ (\beta-2s)}{16\pi \beta}\left(1 - \frac\beta
2\right)( l^{2}\Delta_g)R+\dots,\\&
\tilde\varphi_c= -\frac{  (\beta-2s)}{4}\left(1 - \frac\beta
2\right)l^2R+\dots.  \end{align} 
To recover the value of the action \eq{effective_action} from this solution, we note that the first term in \eq{effective_action} does not contribute to the leading orders, while the  second term yields a constant metric independent value.

Thus, up to a  constant, the classical value of the action $\Gamma[\varphi_{c}] = \Gamma_{c}[g]$, where $ \Gamma_{c}[g]$  is the contribution of  the background charge  (\ref{background_action}-\ref{3.17}).

\subsubsection*{Mixed and Gravitational Anomaly Combined}
 In summary, the  three leading terms of the expansion of the generating functional are \begin{align}\la{13.25}
\log\mathcal{Z}[g]=-\Gamma_{c}[g]+  \frac  12\log{\rm Det}\left(-\Delta_g\right)+O(l^{2}).
\end{align}
This   is equivalent to the result \eq{logZ} obtained through the iteration of the Ward identity.

With the generating functional in hand, we compute the expectation value of the density using \eq{3.7}
\begin{align}\la{4.18}
&\langle\rho\rangle=\rho_c+\frac{1}{8\pi} \frac{1}{12}(l^2\Delta_g)R + ...
\end{align}
where the first term is the  contribution of the background charge and the last term comes from fluctuations governed by the gravitational anomaly.

We observe that the uniform part of the density's response to curvature  \(\eta\)
\eq{response_R} is determined by the background charge. It is a classical phenomenon, which 
also appears in superfluid helium and classical fluids \cite{AW}.  This response is often referred to as the mixed  anomaly 
since it mixes gravitational and electromagnetic responses.

The fluctuation term beyond  the classical density is the response
to curvature. It is a pure  quantum
 effect, referred to as the gravitational anomaly.
 
 In the case of a surface of constant curvature, $\Gamma_c[g]$ consists of a super-extensive $O(N^{2})$ term, arising from the self-interaction energy of the background charge. The subleading contribution is proportional to the number of particles $ O(N)$, and recieves corrections from the fluctuations. The spectral determinant in this case is known to have two relevant contributions (see e.g., \cite{Alvarez})
 \be\la{PPP} \log{\rm Det}\left(-\Delta_g\right)=-\frac 1{4\pi}\frac{V}{ \epsilon^{2}}-\frac \chi {6}\log \frac{V}{ \epsilon^{2}},\ee
 where $\epsilon$ is the length of the   short-distance cut-off. Naturally, it is proportional to the magnetic length $\epsilon\sim l$. As a result, the first term in \eq{PPP} is proportional to $N$, while the second is universal and equal to $-\frac \chi {6}\log  N$. This gives the result previously conjectured by Jancovici et al.
\cite{Jancovici,Jancovici1} for the free energy of the 2D Coulomb gas (a problem of statistical mechanics in which the Boltzmann weight is equal to the probability density of  the Laughlin states). Jancovici et al. conjectured  that the free energy of the Coulomb plasma on a surfaces with a constant curvature receives a universal finite size correction which depends only on the Euler characteristic, implying 
\be\la{J}  -\log\mathcal{Z}[g]= {\rm const}\cdot N^{2} + {\rm const} \cdot N -  \frac \chi {12}\log  N
\ee
The first constant comes from the background-background interaction $\frac{\pi \beta }{2V^{2}} \int K dV$, and is typically moved to the LHS and included in the definition of the free energy. The constant in front of $N$ is only accessible numerically for general $\beta$ \cite{Forrester1}. An important aspect of  the next correction is that it comes with the opposite sign compared to the finite-size corrections in conformal field theory (see, e.g., \cite{Cardy}). Consequently the Casimir forces act in the opposite direction. They are repulsive. The origin of this is the flipped sign due to the contribution of the measure in \eq{4.101}, discussed above. See more about this in \ref{HH}.

 \section{  Laughlin
States in General Coordinates }\la{B} Various terms of the   generating
functional  come together nicely when we write the wave function  in general coordinates. In order to avoid unnecessary complications, we consider the amplitude of the Laughlin state and assume spin $s=0$.  It will also be convenient to work with the magnetic flux $N_{\phi} = V/2\pi l^{2}$.
Recall that in complex coordinates, the modulus of the wave function is
  \begin{align}\la{11.9}|\Psi|^2=\mathcal{Z}^{-1}[g] \prod_{i<j}^N|z_i-z_j|^{2\beta}
e^{-\pi N_\phi\sum_{i=1}^N\frac{1}{V} K(\xi_i)}.
\end{align} 
  In order to pass to general coordinates we express the entries in this
formula in terms of the Green function of the Laplace-Beltrami operator and
use the explicit form of the generating functional we computed in \eq{logZ}. 
The transformation formulas of the \ref{G} suffice.  We set the reference metric $g_0$ to be the metric of a uniform sphere. 

Summing over the coordinates in equation \eq{Ga} and using the formula $N_{\phi} = \beta (N -1 )$, we obtain the identity {  
\begin{align}\la{id1}
\beta \sum_{i\ne j} \log |\xi_{i} \!\!-\!\! \xi_{j}| -\pi N_{\phi}\sum_{i} \frac 1 VK(\xi_{i}) = - 2\pi \beta \sum_{i \ne j} G(\xi_{i}, \xi_{j}) - 2\pi N N_{\phi} \left(A^{(2)}[g, g_{0}] + \frac{1}{4\pi}\right).
\end{align}
}
%
%
Now, employing equations (\ref{Gb} - \ref{Gd}) yields the identity 
\begin{align}\la{id2}
- N_{\phi}\left(\frac 14A^{(1)}[g,g_0] +\pi\chi A^{(2)}[g,g_0]\right)&+ \frac{1}{8\pi} \left( \frac{1}{12} - \frac{\beta}{4}\right) A^{(0)}[g,g_0] =\\&= \frac{\beta}{2} \sum_{i} \left[ \int G(\xi_{i}, \xi') R(\xi') dV' - {  4\pi G^{R}(\xi_{i}) - 1 }\right]
  \nn\\&- \frac{\beta}{32\pi} \int R(\xi) G(\xi, \xi') R(\xi') dV dV'+\frac{1}{96\pi} A^{(0)}[g,g_0]. \nn
\end{align}

Plugging  \eq{id1} into \eq{11.9} and using  \eq{id2} we observe that first three leading terms in the  normalization factor $\mathcal{Z}[g]$ we computed \eq{logZ} almost cancel, and 
\begin{align*}\log |\Psi|^{2} &= -2\pi \beta \left(\sum_{i\neq j} G(\xi_i,\xi_j)+\sum_iG^R(\xi_i)\right) + \frac{\beta}{2} \sum_i\int
G(\xi_i,\xi')RdV'\\&-\frac{\beta}{32\pi}\int
R(\xi)G(\xi,\xi')R(\xi')dVdV'  +\frac{1}{96\pi} A^{(0)}[g,g_0]  + O(N_{\phi}^{-1}),
\end{align*}
{  up to metric-independent constants.} 
The remaining part $\frac{1}{96\pi} A^{(0)}$  is the Polyakov's functional which is equal to the square root of the spectral determinant  of the Laplace-Beltrami operator \eq{P}.  Thus the normalized wave function \eq{11.9} reads
\begin{align}|\Psi|^2=&C\left[{\rm Det}(-\Delta_g)\right]^{-1/2}\exp{\left(-\frac{\beta}{32\pi}\int
R(\xi)G(\xi,\xi')R(\xi')dVdV'\right)}\times\nn\\&\exp{\left(-2\pi\beta\sum_i\left[\sum_{j\neq i} G(\xi_i,\xi_j)+G^R(\xi_i)- \frac 1{4\pi}\int
G(\xi_i,\xi')R(\xi')dV'\right]\right)},\la{13.4}
\end{align}
where the factor $C$ in the leading order in $1/N_\phi$ does not depend on the metric. It depends only on the volume and the filling fraction.

Thus, quite remarkably,  all three geometric  terms of the generating  functional have been completely absorbed through the transformation formulas for the Green function. The normalization factor $C$ of the wave function written in general coordinates does not depend on the metric. 

In this representation, the spectral determinant solely represents the gravitational anomaly. The Laughlin wave function in general coordinates was also recently studied in \cite{KlevtsovGFF}.

\section{ Laughlin State as a String of Vertex Operators of an   Auxiliary Gaussian Field}\la{14SS}
It has been noticed long ago that the Laughlin state  (for a finite $N$)  could  be represented by the Gaussian Free Field Integral. However,  in the flat space such representation  is ambiguous. The essential ingredient of the Gaussian Free Field is the background charge which to the best of our knowledge have not been determined correctly.   
 This issue has been resolved in the recent note by S. Klevtsov \cite {KlevtsovGFF}, who noticed that the wave-function \eq{13.4} has a representation through the Gaussian integral in the curved space 
\begin{align}|\Psi|^2= C \int \exp{\left(i\sum_i\sqrt \beta\, \Phi(\xi_i)\right)}\exp{\left(-\frac 1{4\pi }\int \left[(\nabla\Phi)^2+i\sqrt\beta R\,\Phi\right] dV\right)}D\Phi.\nn
\end{align}
 Using this, Ref.\cite{KlevtsovGFF} independently derives the generating functional \eq{logZ} as the gravitational effective action of the free field theory. In this integral the  auxiliary field \(\Phi\) is defined such that it has no zero modes \(\int\Phi \,dV=0\).
The essential term in this formula is the background charge - the coefficient in front of the  term  $i(\sqrt\beta/4\pi)\int R\Phi $.  In the flat space it vanishes  in the bulk but its reflection  is found on the boundary in the form of the overshoot \cite{CFTW}. In the case of the spin state, the background charge changes by $i(\beta-2s)/(4\pi\sqrt\beta)\int R\Phi $.

In the nomenclature of the Gaussian Field this formula determines the background charge equal to  \(\alpha_0=(\beta-2s)/\sqrt{8\beta}\), such that the central charge of the Gaussian field is $1-24\alpha_0^2=1-3(\beta-2s)^2/\beta$.    The charge of the  vertex operator representing electrons \(e^{i\sqrt \beta\, \Phi(\xi_i)}\)  is \(\alpha=\sqrt {\beta/2}\), and its  holomorphic dimension  $h_\alpha=\alpha(\alpha-2\alpha_0)$  and conformal spin become equal to $s$.


\section{Summary and Discussion}
In this paper, we have presented a detailed account of the geometric response of FQH states. We showed how the transport coefficients of the FQH states are determined by the response of the  ground state of the system to the variation of the spatial geometry. We introduced a notion of the generating functionals which encodes all the properties of states contributing to the transport properties.

A full-fledged response of FQH states to geometry may be complicated and, although mathematically interesting,  may depend on non-universal details of the system. However, some features of the response are  universal and controlled only by global characteristics of the state. In the  paper we identified these features. They are controlled by the anomalous terms of the action of general  diffeomorphisms. These terms are the best seen in $1/N_\phi$  expansion of the generating functional. The three first terms  of this expansion represents the gauge anomaly, mixed anomaly and gravitational anomaly.  Each anomaly controls its own transport coefficient. The gauge anomaly controls the fractionally quantized Hall conductance. The mixed anomaly controls the odd (or anomalous) viscosity. We believe that the gravitational anomaly controls the thermal transport, although a detailed understanding of thermal transport remains an open problem for these states. It is not presented in  this paper.

We focused on  Laughlin states and  presented a conjecture for the Pfaffian or parafermion states. A detailed understanding of the structure of the Ward identity for these states remains an open problem. Our  analysis  can be naturally extended to the study of multi-component states. We did not detail it here.

The major ingredient of the geometric response is 
the short distance behavior of the two-point functions. We showed that the short distance behavior is  influenced by the gravitational anomaly. Extending the expansion of the density to higher orders would require a deeper understanding of this short distance behavior. This is understood for the IQH problem, but remains open for the FQH case.

\section*{Acknowledgments}\noindent We thank A.G. Abanov, A. Cappelli, P.J. Forrester, A. Gromov, I. Gruzberg,  
S. Klevtsov, A. Zabrodin and S. Zelditch for inputs at different
stages of this work. We thank D. T. Son for sharing with us Ref.\cite{SonCourtesy}, which gives an alternative derivation of \eq{projected_s}. The work  was supported by NSF DMS-1206648, DMS-1156656,
DMR-MRSEC 0820054,  John Templeton Foundation and PVE grant from the CNPq-Brazil Science Without Boarders Program. P.\ W. thanks  
 IIP-UFRN for the hospitality where this paper was completed.
\begin{appendix}
\section{Parafermionic FQH Model States} \la{A}
A class of LLL wave functions introduced in Ref.\cite{Read99}, known as parafermion  states, are characterized by two integers $k$ and $M$,
\begin{align}\la{parafermion_wf}
F(z_1,...,z_N)= \mathcal{F}_{k}(z_{1}, ..., z_{N}) \prod_{i < j} (z_{i} - z_{j})^{M + 2/k}.
\end{align}
The  function $\mathcal{F}_{k}$ diverges  as two particles coincide as
\begin{align}\la{Fk_ope}
 \mathcal{F}_{k}(z_{1}, z_{2}, ...., z_{N}) |_{z_1\to z_2}\sim (z_{1} - z_{2})^{- 2/k} .
\end{align}
The divergency  cancels the fractional  power of the second factor such that together $F$ is a polynomial with  degrees of the zero at $z_1\to z_2$ equal to  $M$. 

This function is also {\red a} primary under M\"obius transformations \eq{23}. Let us call its dimension   $ h_{k}$. 
Thus, the wave function has total dimension
\begin{align}
-2h_{N} & = \left(M + \frac{2}{k}\right) (N-1)  - 2h_{k}.
\end{align}
The filling fraction and shift can be read off from this relation by comparing with Eq.\eq{2.23}. They are
\begin{align}\la{parafermion_shift_filling}
\nu = \frac{1}{M + 2/k}, \qquad\mathcal{S} = \nu^{-1} + 2h_{k}.
\end{align}
For odd $M$, these states are antisymmetric and thus describe fermions. 
For even $k$, they describe electrons at even denominator filling fractions. The requirement of integer valued shift requires that $h_{k} = p/2 - 1/k$ for integer $p$. The parafermionic  states correspond to $p = 2$, which implies $\mathcal{S} = M + 2$. The Pfaffian state at half filling occurs at $k = 2$.

Parafermion states
are associated to the affine   \(\widehat{sl}_k(2)\) algebra \cite{Read99}. Other states associated 
with affine groups of higher rank, such as  multi-component  states, are not considered here.

\section{Response to Curvature and Odd  Viscosity}\la{B1}
The response of the density to the curvature is intrinsically related to the
odd viscosity, introduced in \cite{Seiler1995}. We will refer to it as the anomalous viscosity. We briefly recall its definition.

Let \(\Pi_{ab}\) with \((a,b=x,y)\) be the momentum flux tensor of the  QH fluid, and \(v=(v_x,v_y)\) the velocity of the flow. In complex coordinates, \(\Pi=\Pi_{xx}-\Pi_{yy}-2\ii \Pi_{xy}\)
is the holomorphic component of the momentum flux tensor and \(\vv=v_x-\ii v_y\) is the complex velocity of the fluid. Then in the linear approximation and leading order in
gradients
\begin{align}\la{set}
\Pi=2 \ii  \eta^{(A) } \rho\p\vv.
\end{align} 
 The coefficient \(  \eta^{(A) }\) is the kinematic anomalous viscosity.  We further set the effective mass $m^{*} = 1$.

It is instructive to take  a \(\bar \p\) derivative of this equation
and use the  conservation law of the momentum stress tensor kept at the
steady state. The conservation law in complex coordinates reads \(\p\Theta+\bar\p\Pi=0\),
where \(\Theta=\Pi_{xx}+\Pi_{yy}\) is the trace of the tensor.
For an incompressible flow $(\nabla \cdot v = 0)$ it becomes 
\begin{align}\la{2.30}
\Theta=  -\eta^{(A) } \rho(\nabla\times v).
\end{align}
On the other hand the  reciprocal relation for kinetic coefficients states that \begin{align}
\frac{\delta\Theta(\xi)}{\delta\mu({\xi'})}=\frac{\delta\rho(\xi)}{2\delta
\sigma({\xi'})},
\end{align} 
where \(2\delta \sigma=\delta\log\sqrt g\)  is a  variation of the conformal factor of the metric and  \(\delta\mu\)
is a variation of the  chemical potential. This can be rewritten in terms of the density response to curvature as 
\begin{align}\la{2.30a}
-\frac{\delta \Theta(\xi)}{ \delta (\nabla^{2} \mu(\xi'))} = \frac{\delta \rho(\xi)}{ \delta R(\xi')}.
\end{align}
Treating the gradient of the chemical potential as an electric field \(\nabla\delta
\mu=-e\delta E\) and using the relation for the Hall current \(\rho\delta v_{i}=
(e\nu/h)\epsilon^{ij}\delta E_{j}\), for a uniform density $\rho$ we can write \eq{2.30} as $\Theta = -(\nu \eta^{(A)}/h) \nabla^{2} \mu$. Using this in \eq{2.30a} leads to the relation between the anomalous viscosity and the density response to the curvature.
\begin{align}\la{4.7}
\eta^{(A)} = \hbar \eta(k)|_{k=0}.
\end{align}
 Eq \eq{eta} yields \begin{align}
\eta^{(A)}=\frac\hbar 4 (\mathcal{S}-2s).
\end{align}The  formula relating the anomalous viscosity to the shift  was obtained in \cite{Read2009}. 

Deriving this formula we assumed a uniform density. This is justified
only in the leading order in gradients, and implies the formula \eq{4.7} is valid only  for the uniform part of the anomalous
viscosity.

The connection between the anomalous viscosity and the density response to curvature
is also observed in the 2d hydrodynamics of vortex flows \cite{Wiegmann2013}.
 Following \cite{PW13}, particles in the LLL can be viewed  as  vortices in an incompressible superfluid.

\section{Saturation of Dilatation Sum Rule}\la{V}
In this Appendix, we show that the dilatation  sum rule \eq{3.9} is saturated by the leading order density \eq{1}. We include the spin. 

 The dilatation sum rule reads
\begin{align}\la{dil}
\int z \partial_{z} W \langle \rho \rangle dV = - N \left(1- h_{N}\right),
\end{align}
 It is convenient to write the RHS of \eq{dil} in terms of $N_{\phi}$. Setting $ - 2h_{N} = N_{\phi} - 2s$ according to Eq.\eq{3}, and using $N = \nu \left(N_{\phi} + \mathcal{S} - 2s\right)$ according to \eq{shift-spin} for $\chi = 2$, the sum rule reads
\begin{align}\la{dil2}
\int z \partial_{z} W \langle \rho \rangle dV= - \frac{\nu}{2} \left( N_{\phi} + \mathcal{S} - 2s\right)\left(N_{\phi}-2s + 2\right),
\end{align}

The density in this formula is given by \eq{density_spin}, 
which we write in the form \begin{align}\la{7.41}
\langle \rho \rangle  = 
\frac{\nu N_{\phi}}{V} + \frac{\nu  (\mathcal{S} - 2s)}{8\pi} R+ O(N_{\phi}^{-1})
\end{align}
For the source potential we use $W = - \pi N_{\phi} K/ V +  (1 - s) \log \sqrt{g}$ (recall that $N_{\phi} = V/2\pi l^{2}$).

We would like to show that the sum rules are saturated by the first two terms in \eq{7.41}. The last term there and other  terms in the curvature  expansion do not contribute. 

 In order to evaluate the LHS of the sum rule explicitly, we make use of the  integrals
\begin{align}
&\int z \partial KdV = V^{2}/(2\pi),\quad 
\int z \partial \log \sqrt{g} dV  = - V,\nn\\&
\int z \, \partial K \, R \, dV  = 4V,\quad
\int z \, \partial \log \sqrt{g} \, R dV  = - 8\pi.\nn
\end{align}
which follow 
from  the asymptotic behaviors $K \to (V/\pi) \log |z|^{2}$ and $\log \sqrt{g} \to - 2 \log |z|^{2}$ at the marked point $z_0=\infty$. As an example, 
\begin{align}
\int z \partial K \, \sqrt{g}\, d^2\xi = \int z \partial K \partial \bar{\partial} K d^{2} \xi 
= \frac{1}{2}\int \bar{\partial} \left( z (\partial K)^{2}\right) d^{2} \xi = - \frac{i}{4} \oint_\infty dz \, z (\partial K)^{2} = \frac{V^{2}}{2\pi}.\nn
\end{align}
The third step  requires use of Green's theorem. With the contour taken around the marked point, we can make use of the asymptote and simply evaluate the contour integral. The other integral identities are evaluated in a similar manner. 

Equipped with this, a straightforward evaluation of the LHS of \eq{dil} with only the two first orders in the density expansion is seen to yield the RHS. Thus, the leading order density completely saturates the dilatation sum rule. A similar calculation shows that other two sum rules in Eq.\eq{3.9} vanish identically.

\section{Short Distance Behavior}\la{shortd} In this Appendix we present additional arguments supporting the short distance regularization of Sec.\ref{UV_reg}.  

We can write the density for $N$ particles as the expectation value of the vertex operator {  \(e^{-\varphi(\xi)}=\prod_{i=1}^{N-1}|z-z_i|^{2\beta}e^{ - \frac{\pi \beta}{V}K(\xi_{i})}\) }over
the ensemble of \(N-1\) particles
with the magnetic {  flux $N_{\phi} - \beta$} 
\begin{align}
\langle \rho(\xi) \rangle_{N} =N \int |\Psi(\xi,\xi_2,\dots,\xi_N)|^2 dV_2\dots dV_N= N \mathcal{Z}_N^{-1} \mathcal{Z}_{N-1}e^{-\frac \pi
V N_\phi K(\xi) } \langle e^{- \varphi } \rangle_{N-1}\nn.
\end{align}
Employing the identity
\begin{align}\la{6.9}
\langle e^{\mathcal{O}} \rangle = \exp \left( \sum_{k = 1}^{\infty} \frac{\langle \mathcal{O}^{k} \rangle_{c}}{k!}\right)
\end{align}
we obtain the expansion   \begin{align}
 \log \langle \rho \rangle_{N} = \log (N \mathcal{Z}_{N-1}/\mathcal{Z}_N)
-\frac \pi
V N_\phi K- \langle  \varphi \rangle_{N-1} + \frac{1}{2} \langle \varphi^{2}\rangle_{c, N-1}  +...
\end{align} 
 Taking the derivative of this expression we substitute it into
the  Ward identity \eq{loop_diff1} 
\begin{align}
\frac{1}{\pi \beta} \frac{1}{ \langle \rho \rangle}\nabla_{ \bar z} \langle (\partial \varphi)^{2} \rangle_{c} &= \p\left[- {\beta} \log (\sqrt
g \langle \rho \rangle  )- \frac{2\pi}V\beta  K  +   \langle \varphi^{2} \rangle_{c}\right]+\dots
\end{align}
In deriving this formula we took into account the relation $\langle  \varphi \rangle_N-\langle  \varphi \rangle_{N-1}\approx -\frac \pi V\beta K$.

Let us compare orders in this equation. The LHS is $O(l^{2})$, while the first two terms in the RHS are
of the lower order. They must cancel. Thus, in the leading order this formula matches the regularization \eq{phi_green1}. 
 \subsubsection*{Pair Correlation Function} Further support for the short distance regularization
comes from evaluation of the pair distribution function \begin{align}\nn
  {\rm g}(\xi, \xi') = \frac{1}{\langle \rho(\xi) \rangle \langle \rho (\xi')\rangle}\Big\langle \sum_{i \ne j}\frac 1{\sqrt {g(\xi)}} \delta^{(2)}(\xi - \xi_{i}) \frac 1{\sqrt {g(\xi')}}\delta^{(2)}(\xi' - \xi_{j})\Big \rangle. \end{align}
This function is closely connected with the structure factor \eq{3.11} through the two-point density correlation function  $\langle \rho(\xi) \rho(\xi')\rangle_{c}= \langle\rho(\xi)\rangle \langle\rho(\xi')\rangle (g(\xi,\xi')-1) + \frac 1{\sqrt g}\delta^{(2)}(\xi-\xi') \langle \rho(\xi) \rangle$.

At large separation $d(\xi,\xi') >> l$, ${\rm g}(\xi, \xi') \to 1\).  However,
at small separation it vanishes as \({\rm\ g}\sim |\xi-\xi'|^{2\beta}\). 
 This function must not depend on the choice of coordinates on the surface.
Therefore we expect that the  short distance behavior to be \begin{align}\la{g_geo}
 {\rm g}(\xi, \xi') \sim [d(\xi,\xi')]^{2\beta},
\end{align}where \(d(\xi,\xi')\) is the geodesic distance. 

Let us reproduce this property employing the identity  \eq{6.9}. First we write the pair distribution function  as \begin{align}
\langle \rho(\xi) \rangle \langle \rho(\xi')\rangle {\rm g}(\xi, \xi')= \frac{N!} {\mathcal{Z}_{ N}}\frac{\mathcal{Z}_{ N-2}}{(N-2)!}|\xi-\xi'|^{2\beta}e^{-\frac
1{2l^2}[K(\xi)+K(\xi')]} \langle e^{- \varphi(\xi) - \varphi(\xi')}
\rangle_{N-2}.\nn
\end{align}
To leading order the relevant term is  \(
 {\rm g}(\xi, \xi')\sim   |\xi- \xi'|^{2\beta} e^{ \langle \varphi(\xi) \varphi(\xi') \rangle_{c}}\).
At large separation of points the pair correlation tends to $1$  if  \(\langle \varphi(\xi) \varphi(\xi') \rangle_{c}|_{\xi\to\xi'}\sim -2 \beta
\log |\xi - \xi'|\), which is another statement of \eq{phi_green}. Agreement with  the covariant  behavior at small separation \eq{g_geo} requires 
\(
\langle \varphi(\xi) \varphi(\xi') \rangle_{c}\sim -2 \beta\left(\log |\xi - \xi'|+ \log d(\xi,\xi')\right)|_{\xi\to\xi'} \approx 4\pi \beta G^R
\) as in \eq{green_reg}.

\section{ Quasi-holes }\la{Q1}
In this Appendix we demonstrate how the techniques developed in the paper
can be used to obtain characteristics of the quasi-holes state:  the charge, the dipole moment, conformal dimension, and statistics. We comment that the gravitational anomaly does not
affect these characteristics. For clarity, we consider  the states with $s=0$. 

 Introduced by Laughlin \cite{Laughlin1983}, the $n$-quasi-hole state on a compact surface can be written
\begin{align}\la{9.11}
\Psi(\{w\},\{\xi\})  =\mathcal{Z}[\{w\}]^{-1/2}\prod_{i = 1}^{N}\prod_{k= 1}^{n} (z_{i} - w_k)^{a_k} e^{ - \frac{\pi}{2V} a_{k} K(\xi_{i})}\Psi(\{\xi\}).\end{align}
The set \(\{w\}=w_1,\dots,w_n\) are holomorphic coordinates of quasi-holes, the set of integers  \(a_1,\dots,a_n\) are the charges of quasi holes, $\Psi(\{\xi\})=\mathcal{Z}^{-1/2}F(\{z\})
\prod_{i} e^{Q(\xi_{i})/2}$  is the wave function of the ground state \eq{LLL_wf} is the generating functional   which  depends on coordinates and charges of the quasi-holes. In that state the total magnetic flux to accommodate quasi-holes of charge $a_k$ at fixed particle number $N$ is  $ N_{\phi} +\sum_k a_{k}$, where $N_\phi$ retains the relation \eq {shift-spin} \begin{align}\la{E3} N=\nu N_{\phi} + 1.\end{align} 
The factor  $\exp [- \frac{\pi}{V} a_{k} \sum_{i} K(\xi_{i})]$ in \eq{9.11}  reflects the value  of the total flux.  

The quasi-hole state is the coherent state  
obtained by a change of \(Q\) in  \eq{LLL_wf} 
by $\frac 12 Q\to \frac 12 Q+  \sum_{i = 1}^{N}\sum_{k= 1}^{n} a_k[\log(z_{i} - w_k) - \frac{\pi}{2V}K(\xi_{i})]$. 

The normalization factor in \eq{9.1} is the generating functional of the state
\begin{align}\mathcal{Z}[\{w\}]= \int \prod_{i = 1}^{N}\prod_{k=1}^n |z_{i} - w_k|^{2a_k} e^{ - \frac{\pi}{V} a_{k} K(\xi_{i})}dP,\end{align} 
Using the definition of the  field $\varphi$ \eq{4.4} we write it in terms  the vertex operator 
\begin{align}\la{E44}
\mathcal{Z}[\{w\}]=\Big\langle
e^{ - \nu \sum_{k } a_k\varphi(w_{k}) }\Big\rangle.\end{align}
whose   expectation value taken over the state  without quasi-holes. 
\paragraph{Evaluation of the generating functional} 

With the  help of \eq{6.9} we obtain the expansion (keeping only terms depending on coordinates $w_{i}$) \begin{align}\la{logZqh}
\log \mathcal{Z}[\{w\}] &= -\nu\! \sum_k \!\!a_k\langle\varphi(w_k)\rangle\!+\!\frac{\nu^2}{2}\sum_{k}a_k^2 \langle \varphi(w_{k})^{2} \rangle_{c}\! +\! \frac{\nu^2} 2\sum_{k
\neq l } \!a_ka_l\langle \varphi(w_{k})\varphi(w_{l}) \rangle_{c}\!+\!\dots
\end{align}
 We use
 \eq{phi_green}
\(\langle \varphi(\xi) \varphi(\xi') \rangle_{c}= 4\pi\beta G(\xi,\xi')\), and \eq{phi_green1} \(\langle \varphi^{2} \rangle_{c} = 4\pi\beta G^R\) to evaluate the  leading orders

\begin{align}\la{E5}
\log \mathcal{Z}[ \{w\}]  =  \nu\sum_k a_k\left(- \<\varphi(w_k)\>  + 2\pi a_kG^R(w_k)  +2\pi \sum_{l\neq k} a_lG(w_k,w_l)\right).
\end{align}

\paragraph{Adiabatic phase, conformal dimension, and mutual statistics}

Consider an adiabatic motion of a chosen quasi-hole, say \(w_1\), along a closed path \(\mathcal{C}\). Under
this motion the  state  acquires an adiabatic  phase \(\gamma_\mathcal{C}\). 
Since the state is a holomorphic function of the position of the quasi-hole, the phase counts the number of zeros of this function encircled by the path. For that reason it can also be written through the normalization factor in \eq{9.11}\begin{align}
\gamma_\mathcal{C}=2 \ii \int\left[ \oint_\mathcal{C}\overline{\Psi}\partial_{w_1}\Psi
dw_1\right]dV_1\dots dV_N&=\ii \oint_\mathcal{C}\partial_{w_1}\log\mathcal{Z
}[\{w\}]dw_1=\nn \\&=-\frac 1 2\int_\mathcal{C}\Delta_{w_1}\log\mathcal{Z
}[\{w\}]dV_{w_1}.\la{D4}
\end{align} 
The last integral goes over the region surrounded by the contour.

We compute $\Delta_{w_1}\log\mathcal{Z
}[\{w\}]$ with the help of the formula \eq{E5} and the relation 
\begin{align}\Delta_g G^R=\frac 2V-\frac R{4\pi},\end{align} 
which follows from \eq{Gb} and valid for genus zero  surfaces
\begin{align}
\frac 12\Delta_{w_1}\log\mathcal{Z}[\{w\}]= 2\pi   a_1\left(\<\rho(w_1)\>+\frac{\nu} V \sum_ka_k -\nu\frac {a_1} {8\pi} R- \nu\sum_{k\neq 1}a_k\frac{1}{\sqrt{g}}\delta^{(2)}(w_1-w_k)\right).\nn
\end{align}
 Since the density is computed in the state for which \eq{E3} holds, it is given by the Eq.  \eq{1}.
Then, the  adiabatic phase \eq{D4}  becomes  the sum of three terms 
  \begin{align}\la{Berryphase}
\gamma_{\mathcal{C}}= -2\pi a_1  \nu  \frac{\Phi_{\mathcal{C}}}{\Phi_{0}} -
h_{a_1}\Omega_\mathcal{C}+2\pi \nu a_1 \sum_{k\in \mathcal{C}}a_k\end{align}
The first term is  the Aharonov-Bohm phase picked up by a particle with charge $-a_1\nu$ encircling a flux $\Phi_{\mathcal{C}}$.
The second term contains the solid angle \(\Omega_\mathcal{C}=\frac
12\int _\mathcal{C}R dV\). The coefficient in front of it is the holomorphic  conformal dimension  of the vertex operator  \eq{E44} representing the quasi-hole 
\be\la{ha}  h_{a}= \frac{\nu a}{2}(\mathcal{S} -  a)  .
\ee
This formula extends the result of Ref.\cite{Li92}, which was for the adiabatic phase of a single quasi-hole ($a = 1$, $s = 0$) on a sphere. 

 The last term in \eq{Berryphase}  counts the charge of  other  quasi-holes  encircled by the pass. 
The quasi-hole $a_{1}$ encircling another quasi-hole with the charge $a_{2}$ picks the phase $2\pi a_1a_{2}\nu$. This is commonly referred to as the (mutual) statistics of the quasi-holes. We comment that often in the literature, in connection to the relation between spin and statistics,  a quasi-hole is referred to as a state with a spin $\frac 12 \nu$ (at  $a=1$), which is just the last term of the holomorphic dimension in \eq{ha}. It is more accurate to say that the state is characterized by its conformal dimension.

 Higher order terms do not contribute to the adiabatic  phase. Also, notice that the effect of the gravitational anomaly does not show up in this formula. The term with $\Delta_{g} R$ in the density integrates to zero on smooth manifolds.

\subsection*{Charge, the second moment and conformal dimension of the quasi-hole} 
A quasi-hole perturbs the electronic density around itself. In the planar case, the density  depends only on the distance $r$  to the quasi-hole.  At very short distance to the position of the puncture, the density   falls as a power law $\sim r^{2a}$ similar to \eq{g_geo}. At large distances the density approaches to a constant  $(N+a\nu)/V$ featuring  a complicated  oscillation in the crossover.
We will show that this complicated pattern  possesses two universal characteristics - a deficit and second moment.

We denote by $\delta\rho=\<\rho_{N,a}\>-\<\rho_{N+\nu a}\>$ the difference in the  densities of two systems with the total flux $N_\phi+a$. The first is the system of interest. It  consists of  $N=\nu N_\phi+1$ electrons and a quasi-hole, with an electron density $\rho_{N,a}$.   The second  system  consists of $N+\nu a=\nu (N_\phi+a)+1$ electrons and no quasi-holes with a density $\rho_{N+\nu a}$. Both states cover the entire surface and their densities approach the same value away from the  quasi-hole. We will show that 
\begin{align}
\int  \delta \rho\,  d^{2} \xi&= - \nu a,\la{D9}\\
\frac 1{2l^2}\int \,r^{2}\delta \rho\,  d^{2} \xi &= h_{a}-\nu a.\la{D10}
\end{align}
Eq.  \eq{D9} suggests that each quasi-hole is a puncture which fractional
  charge deficit  $\nu a$ and is known since Laughlin's early paper \cite{Laughlin1983}. The result for the second moment seems new. The second  moment  describes the fine structure of the core of the quasi-hole and  is controlled by the conformal dimension \eq{ha}.  Curiously, it vanishes for $\nu=1/3$ and $a=1$. The higher moments are  known only for the integer case.  
  
We show two independent methods to obtain this result.  

The first method relies on  the generating functional  \eq{logZqh}. 
The change of electronic density  caused by a quasi-hole located at $w$ obeys the formula similar to \eq{3.7} 
\begin{align}
\left( 1- \frac{1}{2}(l^{2}\Delta_{g,\xi})\right)\delta\rho(w,\xi)= - \frac{1}{2} (l^{2}\Delta_{g,\xi}) \frac{\delta}{\delta \sqrt{g(\xi)}} \log \mathcal{Z}[w].
\end{align}
 Here, $\log \mathcal{Z}[w]$ is given by  \eq{logZqh} for a single quasi-hole of charge $a$ and the magnetic length corresponds to the total flux $N_\phi+a$. We evaluate this to the leading order as in \eq{E5}, ignoring constants. After taking a variational derivative, we set the metric to be flat, and transform to momentum space. We are left with 
\begin{align}
\left( 1 +\frac{1}{2} (lk)^{2}\right) \delta \rho(k)= - \nu a-\frac{1}{2}h_{a}(lk)^{2} + ...,
\end{align}
which determines  $\delta\rho(k)$ up to the order $k^{2}$
\begin{align}
\<\delta \rho(k)\> = -  \nu a-\frac 12\left(h_{a}-\nu a \right) (lk)^{2} + ....
\end{align}
This formula is equivalent to (\ref{D9},\ref{D10}).

The second method  employs the sum rule of \eq{3.9}. The sum rule for both systems   read   \begin{align}\la{D91}\int z\p_z W\<\rho_{N,a}\>dV&=-\frac 12 N(\beta(N-1)+2+2a),\\
\la{D101}\int z\p_z W\<\rho_{N+\nu a}\>dV&=-\frac 12 (N+\nu a)(\beta(N+\nu a-1)+2),
\end{align}
 Setting it to the flat space form $W=-r^2/2l^2$, and subtracting \eq{D91} from \eq{D101} we obtain \eq{D10}.

%
%
%
%
%
%
%
%
%

\section{Geometric Invariant Functionals}\la{largeN}

In this Appendix, we give more details of the large $N_{\phi}$ expansion, making contact with the existing literature on the Bergman kernel expansion and the Yau-Tian-Donaldson program in K\"ahler geometry. We present the form of the geometric functionals of Sec.(\ref{9.2}) in terms of functions globally defined on the Riemann sphere $2 u=K-K_0$ and $2\sigma=\log \sqrt g-\log \sqrt {g_0}$. The formulas below  can be found in  \cite{Ferrari2012a}. 

We will  keep $\nu$, $\mathcal{S}$, and {  $\kappa$} to describe a general FQH state. For reference, we write the  expansion of the density for large $N_{\phi}$ as
{  \begin{align}\nn
\langle \rho \rangle = \nu \frac{N_{\phi}}{V} + \nu  (\mathcal{S} - 2s)\frac R{8\pi} +\frac 1{2\pi} \frac{V}{N_{\phi}} \left(-\frac{\kappa}{12} + \frac{\nu}{4} (\mathcal{S}-2s)(2 - \mathcal{S})\right)
\Delta_{g}\left( \frac {R }{8\pi}\right)+ O(N_{\phi}^{-2}).
\end{align}
}
Integrating this as we did in Sec \ref{9.2} yields the leading three orders of the generating functional as we  replace $l^{2}$  by $N_{\phi} = V/2\pi l^{2}$ in Eq.\eq{3.7}, {  
\begin{align}\nn
\log \mathcal{Z}[g]/\mathcal{Z}[g_{0}] &= - 2\pi \nu   N_{\phi}^{2} A^{(2)}[g, g_{0}]+ \frac{\nu  (\mathcal{S}-2s)}{4}  N_{\phi} A^{(1)}[g, g_{0}]\\
& + \frac{1}{8\pi} \left( \frac{\nu (\mathcal{S} - 2s)^{2} }{4} + \frac{\kappa}{12}\right)A^{(0)}[g, g_{0}] + ...\nn
\end{align}
}
This follows from  the variations of the functionals
\begin{align}\nn
\delta A^{(2)}[g,g_0] &= \frac{1}{V^{2}} \int \delta u\,  dV, \quad \delta A^{(1)}[g,g_0] = -\frac{1}{V} \int R\,   \delta u \,   dV,\\
\delta A^{(0)}[g,g_0]& =  \int ( \Delta_{g} R ) \delta u \,  dV.\nn
\end{align}
The functionals \eq{genfun1}- \eq{genfun3} can be written in terms of the globally defined functions $2  u = K - K_{0}$, and $2\sigma = \log \sqrt{g}/\sqrt{g_{0}}$, and take the form
\begin{align}
&A^{(2)}[g,g_{0}]  =  \frac{1}{V^{2}}\int \left( u \p\bar\p u + u  \sqrt{g_0} \right) d^2\xi,\la{A2}  \\
& A^{(1)}[g,g_{0}]
 =  \frac{4}{V}\int \left(\sigma e^{2\sigma} - \frac 14 R_{0} u \right) \sqrt{g_{0}}d^2\xi,\la{A1}\\
 & A^{(0)}[g, g_{0}]  = 16 \int \left( - \sigma \p\bar\p \sigma + \frac 14 R_{0}\sigma \sqrt{g_{0}} \right) d^{2} \xi .\la{A0}
\end{align}
The functional $A^{(2)}$ is referred to in the literature as the Aubin-Yau functional. In K\"ahler geometry it is customary to use the functional 
\begin{align} A_{M}^{(1)}&=A^{(1)} + 4\pi\chi {A}^{(2)} \la{mabuchi}\\
&  = \frac{1}{V} \int \left( 4 \sigma e^{2\sigma} + (\bar{R} - R_{0}) u + \frac{1}{4} \bar{R} \, u \Delta_{0} u \right) \sqrt{g_{0}} d^{2} \xi
\end{align}

 This functional is referred as Mabuchi K-energy \cite{Mabuchi}. Similar to Polyakov's Liouville action ${A}^{(0)}$,  the Mabuchi energy is minimized on surfaces  of constant scalar curvature $\bar{R}= \int R dV/V = 4\pi \chi/V$
\begin{align}\nn
\delta A_{M}^{(1)}[g,g_0]=\frac{1}{ V} \int \left( \bar{R} - R\right) \delta u \, dV.
\end{align}
Using $\nu N_{\phi} =  N -  \nu (\mathcal{S} + 2s) \chi/2$ we can also write the generating functional in terms of the Mabuchi energy  
\begin{align}\la{logZmabuchi}
\log \mathcal{Z}[g]/\mathcal{Z}[g_{0}] &= - 2\pi  N\, N_{\phi}  A^{(2)}[ g,g_0]+ \frac{\nu  (\mathcal{S}-2s)}{4}  N_{\phi} A_{M}^{(1)}[ g,g_0]\nn \\
& + \frac{1}{8\pi} \left( \frac{\nu  (\mathcal{S} - 2s)^{2} }{4} + \frac{\kappa}{12}\right)A^{(0)}[ g,g_0].
\end{align}

The functional  \(A^{(0)}\) is  Polyakov's Liouville action known in the theory of 2D quantum gravity \cite{Polyakov1987}. Unlike the higher order terms,
the first three terms cannot be expressed locally through the scalar curvature
$R$. 

The functionals obey the co-cycle condition \eq{cocycle}: $A^{(k)}[g_1,g_3]=A^{(k)}[g_1,g_2]+A^{(k)}[g_2,g_3]$. This condition may not be obvious from (\ref{A2}-\ref{A0}) but are obvious from the form (\ref{genfun1}-\ref{genfun3}).

The three  functionals appear as a result of the transformation properties of Green functions of Laplace-Beltrami operator under a transformation of the metric, as outlined in \ref{G}.

As an exact one-cocycle, the subleading correction to the generating functional must have the structure
\begin{align}\nn
p_{-1} N_{\phi}^{-1} \, \, V \int (R^{2} dV - R_{0}^{2} dV_{0})
\end{align}
For the integer case $\beta = 1$, the asymptotic expansion of the Bergman kernel $\eq{5.19}$ was used in Ref. \cite{Klevtsov2013} to find 
\begin{align}\nn
p_{-1} = -\frac{5}{ 3(16\pi)^{2}}   , \quad (\beta = 1).
\end{align}
Similarly, the next order was found to be
\begin{align}\nn
p_{-2} N_{\phi}^{-2} \, \left(A^{(-2)}[g] - A^{(-2)}[g_{0}]\right),\quad 
p_{-2} =  \frac{1}{180(8\pi)^{3}}, \quad (\beta = 1).
\end{align}
where $A^{(-2)}[g]$ is the local functional of the curvature given by Eq.\eq{A-2}.

\section{Transformation Properties of the Green Function and Geometric Functionals}\la{G} The functionals
(\ref{genfun1}-\ref{genfun3}) that emerged  in the expansion of the generating functional
are invariant
functionals of the manifold. They  do not depend on the choice of
 coordinates.  The functionals have a special role in geometry and could
be  defined by the transformation formulas
with respect to the transformation of the metric. We list these properties
below.   For a more comprehensive discussion see \cite{Ferrari2012a}.

   The transformation formulas involve the Green function of the Laplace-Beltrami
operator
\eq{diff_green1}. They connect geometric objects defined on two cohomological
surfaces \(M\) and \(M_0\) with  
metrics
\(g\) and \(g_0\). The Green function is determined uniquely from the conditions,
\begin{align}\la{def_Green}
- \Delta_{g} G(\xi, \xi') = \frac{1}{\sqrt{g}}\delta^{(2)}(\xi - \xi') - \frac{1}{V}, \quad \int G(\xi, \xi') dV = 0.
\end{align}

The geometric functionals defined in the previous section appear from the transformation properties of the Green function. The Aubin-Yau functional \(A^{(2)}\) appears as 
\begin{align}\la{G6}
&G(\xi,\xi') - G_{0}(\xi, \xi') = \frac 1 {2V} \left(u(\xi)+u(\xi')\right)- A^{(2)}[g,g_0],
\end{align}
where $2u = K - K_{0}$. The formula follows by the requirement that both $G$ and $G_{0}$ satisfy Eq. \eq{def_Green} in terms of the metrics $g$ and $g_{0}$, respectively. Another version of this formula involves
the regularized Green function 
\begin{align}
G^{R}(\xi) = \lim_{\xi \to \xi'} \left( G(\xi, \xi') + \frac{1}{2\pi} \log \frac{d(\xi, \xi')}{l}\right),
\end{align}
where {  $l$ is an arbitrary, metric-independent length scale, a magnetic length in this context.} From this definition, we can extract from \eq{G6} the following relation
 \begin{align}\la{G_reg}
&G^R - G_{0}^{R} = \frac 1{2\pi} \sigma + \frac {1}{ V} u -A^{(2)}[g,g_0],
\end{align}
where $2 \sigma = \log \sqrt{g} - \log \sqrt{g_{0}}$. Integration of the transformation formula with the curvature yields the
transformation formula which involves the Mabuchi functional, 
\begin{align}  \la{11.2}
\int  G(\xi,\xi')R(\xi')dV' - & \int  G_0(\xi,\xi')R_0(\xi')dV_{0}' \\
&= 2 \sigma(\xi) + \frac{1}{2} \bar{R} u(\xi) -\frac{1}{2}A_{M}^{(1)}[g, g_{0}] - 2\pi \chi A^{(2)}[g, g_{0}],
\end{align}
where  $\bar{R} = \frac{1}{V} \int R dV$, and $A_{M}^{(1)} = A^{(1)} + 4\pi\chi  A^{(2)}$ is the Mabuchi K-Energy functional \eq{mabuchi}. Further integration expresses the Liouville action \(A^{(0)}\) in terms of
the Green function 
\begin{align}
\int \int R(\xi)G(\xi,\xi')R(\xi') dV dV' - \int \int R_0(\xi)&G_0(\xi,\xi')R_0(\xi')dV_{0} dV_{0}'\\
&= A^{(0)}[g,g_0]- 4\pi\chi  A_{M}^{(1)}[g,g_0].
\end{align} 
In the case when the reference surface is the uniform sphere,
 {  \begin{align}\nn G_0(\xi,\xi')=-\frac{1}{2\pi}\log\frac{|\xi-\xi'|}{\sqrt{(1+|\xi|^2)(1+|\xi'|^2)}}- \frac{1}{4\pi},\quad
G^R_0=- \frac{1}{4\pi},\quad K_0=\frac V{\pi}\log(1+|\xi|^2),
\end{align}}
the formulas simplify.  The integrals which involve \(G_0\) vanish and the
transformation formula \eq{G6} expresses the Green function in terms of
the K\"ahler potential 
\begin{align}
 &{  G(\xi,\xi') = - \frac{1}{2\pi}\log|\xi-\xi'| + \frac {1} {4V} \left(K(\xi)+K(\xi')\right) - \frac{1}{4\pi} - A^{(2)}[g, g_{0}]},\la{Ga}\\
&{  G^R = \frac 1{4\pi}\log\sqrt g + \frac {1}{ 2V}K- \frac{1}{4\pi} -A^{(2)}[g,g_0],}\la{Gb}\\
\int &{  G(\xi,\xi')R(\xi')dV' = 4\pi \left(G^{R}+\frac{1}{4\pi}\right) - \frac{1}{2} A_{M}^{(1)} [g,g_0]},\la{Gc}\\
\int &R(\xi)G(\xi,\xi')R(\xi') dV dV' =A^{(0)}[g,g_0]- 8\pi A_{M}^{(1)}[g,g_0]. \la{Gd}
\end{align}

\section{Density Functional} \la{HH}

 The structure of the generating functional   $\mathcal Z$  in \eq{gen_fun}  and  the role of the gravitational anomaly are clarified through the density functional, that is the  Legendre transform of $\mathcal Z$ with respect to $W$
\begin{align}
\beta\mathcal{ F}[ n ] =\text{min}_W\left[ \int W  n\, d^{2} \xi - \log \mathcal{Z}[W]\right].
\end{align} 
The density functional has an interpretation in terms of the thermodynamics 2D Dyson gas (or one component 2D plasma).  In this interpretation  the probability density $dP$ of the Laughlin state is seen  as the Boltzmann weight of canonical ensemble of particles on a plane with repulsive 2D Coulomb potential, the function  $W(\xi)$ is a local chemical potential, $n(\xi)$ is the  density of particles, $\beta$ is the inverse temperature, and $- \beta \log \mathcal{Z}$ is a thermodynamic potential. Then the functional $\mathcal{ F}[ n]$ is the  free energy at local equilibrium.

The equilibrium statistical mechanics of the 2D Dyson gas on a plane, uniform sphere  and torus has been extensively studied. See \cite{Forrester_book}  for a comprehensive account. The formulas we obtained extend the results for local equilibrium, where the density slowly varies through the space. 

The Legendre transform can be computed directly from the generating functional \eq{8.3}, but a simpler derivation starts with the Ward identity in the form \eq{loop_diff1}. To  the relevant leading orders we can approximate $n=\<\rho\>\sqrt g$ by $(N/V)\sqrt g$. Within this approximation  the anomalous term \( \frac{1}{2\pi \beta  \langle \rho \rangle}  \nabla_{\bar{z}} \langle ( \partial \varphi)^2 \rangle_{c}
 \)  in \eq{loop_diff1}  by virtue of \eq{6.8a} equals  to  $  \frac{1}{12 \pi }\frac 1{\<\rho\>}\p(\frac 1{\sqrt g} \p\bar\p \log\sqrt g)$, and can also be written as \( \frac{1}{12\pi} \partial \left(\frac{1}{n} \p\bar\p\log
n\right)\).    We obtain the Ward identity in the form which expresses \(W\) through \(n\)
  \begin{align}\la{loop_gas1}
 \partial W =  \partial \langle \varphi \rangle  +  \left(1 - \frac{\beta}{2}\right) \partial
 \log  n  - \frac{1}{12\pi}\partial \left(\frac{1}{n} \partial \bar{\partial} \log
 n\right).
 \end{align}
This form of the Ward identity could be integrated to give
\begin{align}\la{loop_gas2}
 W[n] + W_{0} =\langle \varphi \rangle  +  \left( 1- \frac{\beta}{2}\right)  \log (n  V/N)  -  \frac{1}{12\pi} \frac{1}{n} \p\bar\p \log n ,
\end{align}
where $W_{0}$ is an integration constant. We can determine it for a surface of constant curvature, say a  uniform sphere. In this case, \(W=(\frac 12 N_\phi+1)\log\sqrt {g_0},\;\;n_0=(N/V)\sqrt{g_0}\), \( \langle \varphi \rangle = \frac{1}{2}\beta N \log \sqrt{g_{0}}$, 
and the anomalous term \(-  \frac{1}{12\pi} \frac{1}{n} \p\bar\p \log n=\frac{1}{6 N
} \). This implies $W_{0} =  \frac{1}{6N}$. The LHS \(\beta\mu=W[n]+W_{0}\) is the local chemical potential of the grand canonical ensemble.
Using the
relation \(\mu\delta n= \delta \mathcal{F} \) we  find the  density functional (of the Dyson gas)
\begin{align}\la{8.11}
&\beta\mathcal{F}[n] =-\beta \int  n(\xi) \log\left|\xi-\xi'\right| n(\xi')d^2\xi d^2\xi' +\\ &+  \left(1\!-\!\frac{\beta}{2}\right) \int n\, 
 \log (n )\,d^2\xi - \frac{1}{24\pi} \int \!\log (n V) \,\p\bar\p \log (n) \,d^2\xi.\nn\end{align}
This expression misses metric-independent terms of $O(N)$, which are known exactly for $\beta = 1$ and numerically for certain integer $\beta$ \cite{Forrester_book}. The gravitational anomaly appears as the last term in the density functional.
%
%


\subsubsection*{Gravitational anomaly as a correction to the Boltzmann entropy}
 The last term in \eq{8.11}  can be interpreted as a correction to the  entropy at local equilibrium \(S[n]=-\p\mathcal{F}[n]/\p(1/\beta)\),
 \begin{align} 
S[n]=-\int n
 \log (n ) \,d^2\xi  +\frac{1}{24\pi} \int  \log (n V)\,\p\bar\p \log (n)\,d^2\xi 
\end{align}  
The first term in this expression is the Boltzmann entropy. The last term is the gravitational anomaly. Unlike the Boltzmann entropy, the correction depends on the the modulation of density. This correction is strictly negative. The gravitational anomaly lowers the entropy.  We comment that this formula and also \eq{8.11} misses constant terms (independent of local variations in $n(\xi)$ and the metric) of the order of $O(N)$.

In the case of a  surface  of constant curvature the correction to the entropy equals to $-\frac\chi{12}\log N$. This is the result of Ref. \cite{Jancovici,Jancovici1}, see also \cite{Forrester1} for the numerical support of this term. It shows that at a given volume  the free energy of the Dyson gas on surfaces of constant curvature  receives a universal finite size correction which depends on the genus of the surface.

 In fact, the anomalous term is  the universal part of the  Polyakov's Liouville action. We emphasize that this term comes with the opposite sign to a similar term known in conformal field theory. An alternative  explanation of the sign flip is in the Sec.\ref{4}.

\end{appendix}

\bibliography{long_paper_refs_final_annals_v4}
\end{document}